\documentclass[10pt,letterpaper]{article}
\usepackage{arxiv}
\usepackage[T1]{fontenc}

\usepackage{amsmath}
\usepackage{graphicx}
\ifdefined\pdfminorversion
\fi

\usepackage{algorithmic}
\usepackage{multirow}
\usepackage[table]{xcolor}
\usepackage{colortbl}
\usepackage{verbatim}
\usepackage{acronym}
\usepackage{enumitem}
\usepackage{subfig}
\usepackage{booktabs}
\usepackage{tabularx}
\usepackage{array}
\usepackage{makecell}
\usepackage{placeins}
\usepackage[hidelinks]{hyperref}
\usepackage[numbers,sort&compress]{natbib}
\usepackage{xurl}
\newcommand{\Description}[1]{}

\acrodef{ehmi}[eHMI]{External Human-Machine Interface}
\acrodef{obu}[OBU]{On-board Unit}
\acrodef{rsu}[RSU]{Roadside Unit}
\acrodef{v2i}[V2I]{Vehicle to Infrastructure}
\acrodef{v2v}[V2V]{Vehicle to Vehicle}
\acrodef{v2x}[V2X]{Vehicle to Everything}
\acrodef{v2p}[V2P]{Vehicle to Pedestrian}
\acrodef{v2n}[V2N]{Vehicle to Network}
\acrodef{denm}[DENM]{Decentralized Environmental Notification Message}
\acrodef{cam}[CAM]{Cooperative Awareness Message}
\acrodef{vam}[VAM]{Vulnerable Road User Awareness Message}
\acrodef{vru}[VRU]{Vulnerable Road User}
\acrodef{etsi}[ETSI]{European Telecommunications Standards Institute}
\acrodef{its}[ITS]{Intelligent Transportation System}
\acrodef{c-its}[C-ITS]{Cooperative Intelligent Transportation System}
\acrodef{dds}[DDS]{Data Distribution Service}
\acrodef{ros2}[ROS2]{Robot Operating System 2}
\acrodef{its-g5}[ITS-G5]{Intelligent Transportation System Generation 5}
\acrodef{ar}[AR]{Augmented Reality}
\acrodef{mr}[MR]{Mixed Reality}
\acrodef{c-v2x}[C-V2X]{Cellular - Vehicle To Everything}
\acrodef{vanet}[VANET]{Vehicular Ad-hoc Network}
\acrodef{av}[AV]{Autonomous Vehicle}
\acrodef{atcll}[ATCLL]{Aveiro Tech City Living Lab}
\acrodef{mec}[MEC]{Multi-Access Edge Computing}
\acrodef{5gaa}[5GAA]{5G Automotive Association}
\acrodef{ui}[UI]{User Interface}
\acrodef{ux}[UX]{User Experience}
\acrodef{hmd}[HMD]{Head Mounted Display}
\acrodef{vm}[VM]{Virtual Machine}
\acrodef{e2e}[E2E]{End-to-End}
\acrodef{p2p}[P2P]{Peer-to-Peer}
\acrodef{rtsp}[RTSP]{Real Time Streaming Protocol}
\acrodef{ati}[ATI]{Affinity for Technology Interaction}
\acrodef{sus}[SUS]{System Usability Scale}
\acrodef{asr}[ASR]{Automatic Speech Recognition}
\acrodef{stt}[STT]{Speech-To-Text}
\acrodef{ai}[AI]{Artificial Intelligence}
\acrodef{nlu}[NLU]{Natural Language Understanding}
\acrodef{wer}[WER]{Word Error Rate}
\acrodef{cer}[CER]{Character Error Rate}
\acrodef{vad}[VAD]{Voice Activation Detection}
\acrodef{tts}[TTS]{Text-To-Speech}
\acrodef{dtw}[DTW]{Dynamic Time Warping}
\acrodef{hud}[HUD]{Heads-up Display}
\acrodef{cav}[CAV]{Connected and Automated Vehicles}
\acrodef{fov}[FOV]{Field of View}
\acrodef{xr}[XR]{Extended Reality}
\acrodef{vr}[VR]{Virtual Reality}
\acrodef{hmi}[HMI]{Human-Machine Interface}
\acrodef{fleurs}[FLEURS]{Few-shot Learning Evaluation of Universal Representations of Speech}
\acrodef{its-s}[ITS-S]{ITS-Station}
\acrodef{kpi}[KPI]{Key Performance Indicator}

\definecolor{lightgray}{gray}{0.9}
\definecolor{darkgray}{gray}{0.6}
\newif\ifshownotes
\shownotestrue

\newcommand{\deleted}[1]{}

\usepackage{shellesc}
\newif\ifshowwordcount
\showwordcounttrue
\newcommand{\quickwordcount}[1]{%
  \ShellEscape{%
    texcount -1 -sum -merge -q "#1.tex" > "#1-words.sum"%
  }%
  \IfFileExists{#1-words.sum}
    {\input{#1-words.sum}}
    {\textbf{Word count unavailable}}%
}

\shownotesfalse
\begin{document}

\title{Evaluating Strategies for Multimodal Bidirectional Interactions between Pedestrians and AVs}

% ACM
% Author information is hidden automatically by the anonymous option above.
% Keep the real metadata here so the camera-ready switch is easier later.
%\author{André Clérigo}
%\affiliation{%
%  \institution{Universidade de Aveiro}
%  \city{Aveiro}
%  \country{Portugal}}
%\affiliation{%
%  \institution{Instituto de Telecomunicações}
%  \city{Aveiro}
%  \country{Portugal}}
%\email{andreclerigo@ua.pt}
%\author{Maximilian Schrapel}
%\affiliation{%
%  \institution{Karlsruhe Institute of Technology}
%  \city{Karlsruhe}
%  \country{Germany}}
%\email{maximilian.schrapel@kit.edu}
%\author{Pedro Rito}
%\affiliation{%
%  \institution{Universidade de Aveiro}
%  \city{Aveiro}
%  \country{Portugal}}
%\affiliation{%
%  \institution{Instituto de Telecomunicações}
%  \city{Aveiro}
%  \country{Portugal}}
%\email{pedrorito@ua.pt}
%\author{Susana Sargento}
%\affiliation{%
%  \institution{Universidade de Aveiro}
%  \city{Aveiro}
%  \country{Portugal}}
%\affiliation{%
%  \institution{Instituto de Telecomunicações}
%  \city{Aveiro}
%  \country{Portugal}}
%\email{susana@ua.pt}
%\author{Alexey Vinel}
%\affiliation{%
%  \institution{Karlsruhe Institute of Technology}
%  \city{Karlsruhe}
%  \country{Germany}}
%\affiliation{%
%  \institution{Halmstad University}
%  \city{Halmstad}
%  \country{Sweden}}
%\email{alexey.vinel@kit.edu}

\hypersetup{pdftitle={Evaluating Strategies for Multimodal Bidirectional Interactions between Pedestrians and AVs}, pdfauthor={André Clérigo, Maximilian Schrapel, Pedro Rito, Susana Sargento, Alexey Vinel}}
\author{%
  \href{mailto:andreclerigo@ua.pt}{André Clérigo}\textsuperscript{1,2},\enspace
  \href{mailto:maximilian.schrapel@kit.edu}{Maximilian Schrapel}\textsuperscript{3},\enspace
  \href{mailto:pedrorito@ua.pt}{Pedro Rito}\textsuperscript{1,2},\enspace
  \href{mailto:susana@ua.pt}{Susana Sargento}\textsuperscript{1,2},\enspace
  \href{mailto:alexey.vinel@kit.edu}{Alexey Vinel}\textsuperscript{3,4}\\[4pt]
  \normalfont\textsuperscript{1}Universidade de Aveiro, Aveiro, Portugal
  \enspace\textsuperscript{2}Instituto de Telecomunicações, Aveiro, Portugal\\
  \normalfont\textsuperscript{3}Karlsruhe Institute of Technology, Karlsruhe, Germany
  \enspace\textsuperscript{4}Halmstad University, Halmstad, Sweden
}
\date{}

%\settopmatter{printacmref=false}
%\makeatletter
%\renewcommand\footnotetextauthorsaddresses[1]{}
%\renewcommand\footnotetextcopyrightpermission[1]{}
%\makeatother

\begin{abstract}
Autonomous Vehicles (AVs) share spaces with Vulnerable Road Users (VRUs) in encounters where driver cues are absent. This can leave people uncertain whether the vehicle has noticed them and understood their request. Existing research has studied how AVs communicate with pedestrians through external human--machine interfaces. This work investigates bidirectional multimodal VRU--AV interactions combining Augmented Reality (AR), speech, and gestures. The study follows four stages: literature synthesis of interactions, a survey with 124 participants, interaction loop implementation, and feasibility evaluation. Participants preferred AR cues that made vehicle motion or warnings explicit, high-level voice commands, and conventional gestures. Prototype measurements revealed a trade-off between faster local processing and more accurate infrastructure-assisted processing. The demonstration shows bidirectional interaction using AR wearables and vehicular communications, while leaving decision-making to the AV. Together, the results support treating AV--VRU interaction as a closed loop of intent expression, recognition, vehicle decision, and feedback.
\end{abstract}

%\begin{CCSXML}
%<ccs2012>
%<concept>
%<concept_id>10003120.10003138.10003140</concept_id>
%<concept_desc>Human-centered computing~Ubiquitous and mobile computing systems and tools</concept_desc>
%<concept_significance>500</concept_significance>
%</concept>
%<concept>
%<concept_id>10003120.10003121.10003124.10010392</concept_id>
%<concept_desc>Human-centered computing~Mixed / augmented reality</concept_desc>
%<concept_significance>500</concept_significance>
%</concept>
%<concept>
%<concept_id>10003120.10003121.10011748</concept_id>
%<concept_desc>Human-centered computing~Empirical studies in HCI</concept_desc>
%<concept_significance>500</concept_significance>
%</concept>
%<concept>
%    <concept_id>10010147.10010178.10010179.10010183</concept_id>
%    <concept_desc>Computing methodologies~Speech recognition</concept_desc>
%    <concept_significance>300</concept_significance>
%</concept>
%<concept>
%    <concept_id>10003033.10003079</concept_id>
%    <concept_desc>Networks~Network performance evaluation</concept_desc>
%    <concept_significance>300</concept_significance>
%</concept>
%</ccs2012>

%\ccsdesc[500]{Human-centered computing~Ubiquitous and mobile computing systems and tools}
%\ccsdesc[500]{Human-centered computing~Mixed / augmented reality}
%\ccsdesc[500]{Human-centered computing~Empirical studies in HCI}
%\ccsdesc[300]{Computing methodologies~Speech recognition}
%\ccsdesc[300]{Networks~Network performance evaluation}

\keywords{Augmented reality, vulnerable road users, pedestrian safety, autonomous vehicles, multimodal interaction, speech-to-text, gestures, human-centered interaction design}

\maketitle

\section{Introduction}
\begin{figure*}
  \centering
  \includegraphics[width=\linewidth]{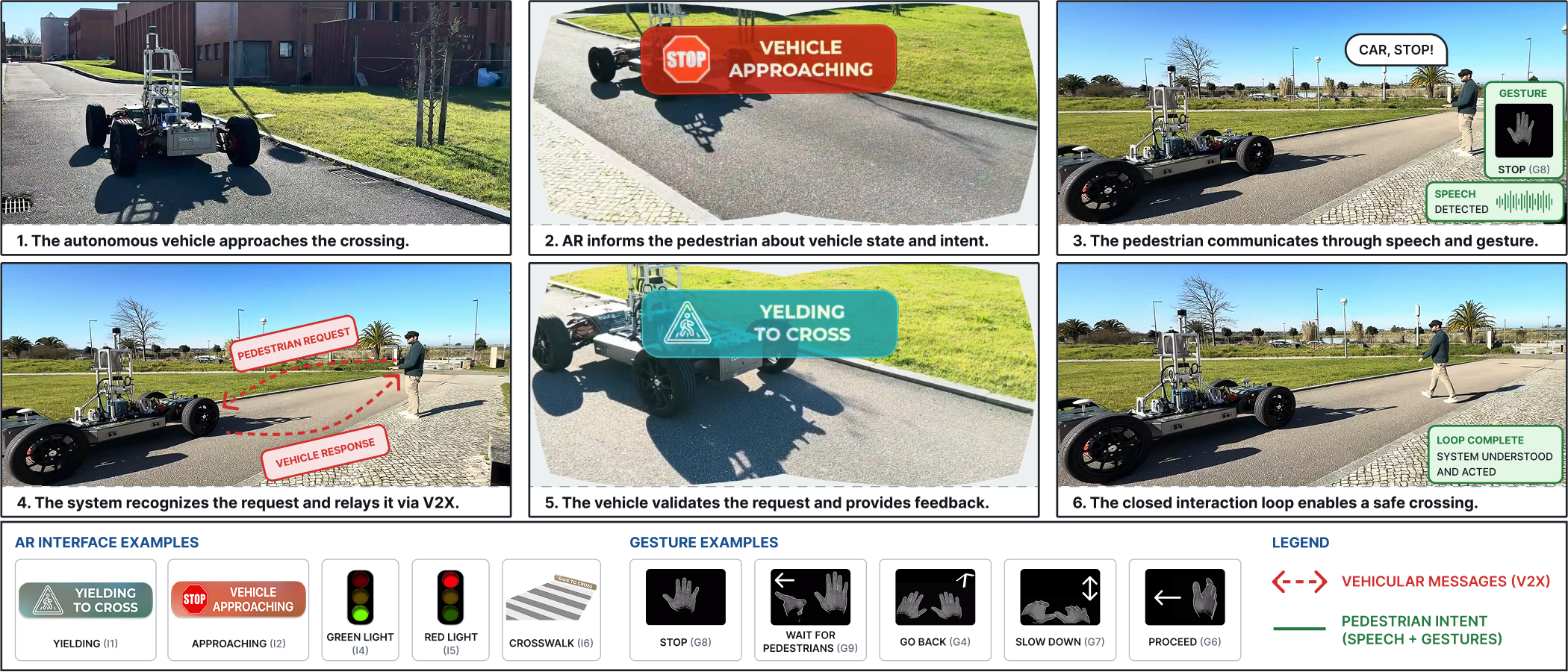}
  \caption{\label{fig:scenario}End-to-end closed-loop interaction between an \acs{ar}-equipped pedestrian and an \acs{av}. The \acs{av} communicates its state and intent through \acs{ar}; the pedestrian responds using speech, gesture, or both; the recognized request is relayed through vehicular communications; and the \acs{av} validates the request before returning explicit \acs{ar} feedback. The lower strip shows representative \acs{ar} interfaces and gesture candidates evaluated in this work.}
  \Description{Visual representation of the use case. A pedestrian wearing an \acs{ar} headset interacting with autonomous traffic through vehicular messages.}
\end{figure*}

The future of urban mobility is trending towards mixed autonomous traffic, with many cities already operating mixed traffic (i.e., traditional, connected, semi-autonomous, and fully \acp{av}). This transition is reflected in public-road testing and deployment programs documented by transport authorities, including the National Highway Traffic Safety Administration \ac{av} TEST initiative\footnote{\url{https://www.nhtsa.gov/automated-vehicle-test-tracking-tool}} in the US and European policy on cooperative, connected, and automated mobility~\cite{european_commission_road_safety_2019}. In the US, commercial robotaxi services such as Waymo already operate fully driverless, although still constrained to operate in a limited number of cities \cite{kusano_waymo_2025}\footnote{\url{https://waymo.com/safety/impact}}.

This transition changes the safety problem from vehicle automation alone to mutual intelligibility in mixed traffic. %In urban environments, safety depends not only on whether
An \ac{av} must detect, classify, and plan around a \ac{vru}; %, but also on whether
surrounding road users must also understand its state and intended behavior, % and whether
while the vehicle infers human intent in ambiguous, negotiated interactions \cite{rasouli_autonomous_2020,rouchitsas2019ehmi}. This is particularly relevant for pedestrians, cyclists, and other \acp{vru}, whose interactions with vehicles often rely on cues such as yielding behavior, driver eye contact, hand gestures, head movements, and speech \cite{rouchitsas2019ehmi,tabone_vulnerable_2021}.

Most research has addressed explicit \ac{av}-to-\acp{vru} communication. \Acp{ehmi}, projections, light bands, displays, and symbolic cues can communicate whether an \ac{av} has detected a pedestrian, is yielding, or intends to continue moving \cite{clamann2017displays,projectionbasedcommunication,vr_av_pedestrian_survey}. These supplement implicit communication through vehicle speed, deceleration, and trajectory, which remains the most common way for road interaction. Explicit bidirectional communication becomes relevant when those cues are ambiguous or when either party needs to express a request or warning. Recent work shows that people may expect their own signals to be recognized and that pedestrian communication includes more than a binary crossing request \cite{colley2021feedbackav,tran2024gesturemapping,chang2024gesture}. A future \ac{av} interaction system should, therefore, support \ac{vru}-initiated input and report whether it was recognized and accepted.

This paper addresses this gap by studying bidirectional multimodal interaction between \ac{ar}-equipped \acp{vru} and \acp{av}. Wearable \ac{ar} presents spatially situated, egocentric, and user-specific feed in the \acp{vru}'s \ac{fov} \cite{tran2022wearablear,aleva2024outdoorar,safearcross}, while the wearable system captures speech and hand movement. Voice and gesture allow the \ac{vru} to express high-level intents, such as asking the \ac{av} to stop, slow down, or proceed. After recognition, a request can reach the vehicle through a direct vehicular link or through cellular or roadside infrastructure, with processing placed locally or in a nearby computing service; this placement may change feedback delay and recognition accuracy. This setup is deliberately interchangeable, working towards the human-facing requirement, which is the explicit feedback to the user that closes the interaction loop.

\autoref{fig:scenario} illustrates the complete bidirectional multimodal interaction sequence. As the \ac{av} approaches, the \ac{ar} headset presents vehicle state or intent; the \ac{vru} responds through speech, gesture, or both; the recognized request is relayed to the vehicle; and the \ac{av} validates it before returning an explicit acknowledgment through \ac{ar}. In the illustrated example, the vehicle accepts the pedestrian's stopping request and communicates that it is yielding, thereby closing the interaction loop. Vehicle motion remains visible throughout and should remain consistent with vehicle response.

This paper makes the following contributions:
\begin{itemize}
    \item We formulate a closed-loop, bidirectional multimodal interaction that restores a channel for human agency while preserving the \ac{av}'s responsibility, and synthesize candidate vocabularies for \ac{ar}, speech, and gesture interactions.
    \item Through an online survey with 124 participants, we provide empirical evidence about which \ac{ar} cues, voice commands, and gestures people perceive as clear and applicable, while identifying ambiguity, semantic overlap, and cultural bias.
    \item We implement the complete interaction loop across a wearable device, speech and gesture recognition, vehicular communication, and an \ac{av}, and demonstrate it in a real outdoor environment. Additionally, we evaluate its technical feasibility through key performance indicators.
    \item We evaluate response delay and speech-recognition error against human-interaction and safety reference points and previous studies.
    \item We derive design implications for closed-loop multimodal interaction between \acp{av} and \acp{vru}, including intent-level commands, optional modalities, uncertainty handling, and fail-safe behavior.
\end{itemize}

The remainder of this paper is organized as follows. \autoref{section:rw} reviews related work. %on \ac{av}--\ac{vru} communication, bidirectional interaction, multimodal input, and wearable \ac{ar}. 
\autoref{section:interactions} defines the interaction goals, candidate modalities, and describes the interaction system. \autoref{section:implementation} describes the architecture of the prototype. \autoref{section:online} presents the online survey, and \autoref{section:results} reports its results and the prototype's feasibility. \autoref{section:discussion} discusses the survey's results and limitations, and \autoref{section:conc} concludes with future work. 

\section{Related Work}
\label{section:rw}
Recent literature on \ac{vru}--\ac{av} interactions has focused primarily on how an \ac{av} communicates explicitly with pedestrians, alongside work on implicit communication through vehicle behavior. Comparatively fewer studies examine how pedestrians or other road users can intentionally communicate back and receive explicit acknowledgments. Systematic reviews show that this body of work remains methodologically focused on \ac{vr} and Wizard-of-Oz experiments~\cite{vr_av_pedestrian_survey,tran2023scoping}. Tran et al. \cite{vr_av_pedestrian_survey}, for example, reviewed 31 \ac{vr} studies on \ac{av}–pedestrian interaction and showed that most relied on controlled simulations.

This type of interaction approach can be categorized as system-initiated from the \ac{vru}'s perspective because it is initiated by the \ac{av} that detects the pedestrian, assesses the situation, and communicates its state or intent. External \acp{hmi}, projected cues, light bands, textual displays, and symbolic interfaces have been explored to communicate yielding, awareness, or right-of-way negotiation \cite{clamann2017displays,howshouldavsinteract,projectionbasedcommunication}. These interfaces differ in visual form and placement, but they share a common assumption that the pedestrian receives and interprets information produced by the vehicle correctly. This assumption becomes problematic, as these interactions are challenged by perception and social factors, as pedestrians must infer whether the communicated intention applies to them or to a nearby person \cite{rasouli2020avsinteract,rouchitsas2019ehmi,kang2025agent}. In contrast, this work explores explicit participant-initiated interaction, where \acp{vru} send a request, acknowledgment, or warning, and receive feedback about whether it was understood. 

The current state-of-the-art has shifted from asking whether an \ac{ehmi} should exist to asking what it should communicate and whether the message remains understandable in complex traffic. Both messages that describe a vehicle's state and messages that describe what it plans to do next can also mislead if they are inconsistent with vehicle motion \cite{eisele2025,Izquierdo2025}. 
Holländer et al.~\cite{hollander2019overtrust}, for example, showed that highly legible or reassuring interfaces can also create overtrust if pedestrians rely on the display without checking whether vehicle behavior is consistent with the message. Scalability further complicates this problem. Single-vehicle crossing concepts may become ambiguous with multiple vehicles, pedestrians, lanes, or shared-space trajectories. Additionally, such scenarios might create information overload in users \cite{colley2023scalability,tran2023scoping,multi_pedestrian_vr_simulator}. Richer forms of spatial information, including intended paths, vehicle connectivity, and shared space design, have also been explored in \cite{intended_path_projection,tran2025connectivity,wang2025placemaking}.

Compared with this mature body of \ac{av}-to-\ac{vru} signaling, \ac{vru}-to-\ac{av} communication remains less developed. This gap matters because pedestrians and cyclists repeatedly signal urgency, hesitation, permission, warnings, and negotiation through gestures, gaze, posture, movement, and speech. Colley et al.~\cite{colley2021feedbackav} showed that pedestrians may expect acknowledgment when they initiate interaction with an \ac{av}, suggesting that bidirectional systems should detect signals and communicate whether they were understood. Gesture-mapping studies further show that pedestrian-to-driver communication includes stopping, proceeding, urgency, warnings, obstacle avoidance, and help-seeking, rather than only crossing intent \cite{tran2024gesturemapping}. The challenge for bidirectional systems is, therefore, to select a small set of inputs that are meaningful, socially plausible, recognizable in real time, and safe enough to be acknowledged by an \ac{av}. Epke et al.~\cite{epke2021gesture} showed that pedestrians could signal through gestures, the vehicle responded through an \ac{ehmi}, and participants preferred the two-way condition. The present work extends that direction by implementing a wearable, multimodal loop and by testing it with an \ac{av}.

Gesture and voice are promising and under-explored candidates for this active direction. \ac{vr} and \ac{mr} studies suggest that familiar gestures can improve interaction efficiency and usability, while poorly chosen gestures can create ambiguity about meaning, target, or authority \cite{gruenefeld2019vroad,chang2024gesture}. Gesture recognition techniques can support rapid prototyping, but they do not remove social and cultural variability in public gestures \cite{vatavu2012pclouds,culturaldronegestures}. Speech can state intents more explicitly than a gesture, yet it is affected by language accents, privacy concerns, noise from road traffic, and overlapping speakers. There is also a higher degree of complexity to encompass open-ended commands that may imply more control than the system can grant \cite{zhang2024externalspeech}. Therefore, multimodal work should add multiple and optional interaction channels to support safer degradation and accessibility of the most viable channel~\cite{dey2024multimodal}. Interaction inside \acp{av} has likewise combined voice and pointing gestures to support spontaneous requests \cite{saurasperez2017voge}. Although that system addressed vehicle occupants rather than external road users, these challenges directly reinforce the closed-loop design adopted here.

Wearables and \ac{ar} devices can support vehicle-initiated feedback and \ac{vru}-initiated input. In the first direction, they can present vehicle state, collision warnings, crossing advice, or spatially registered hazards directly in the \ac{vru}'s \ac{fov} \cite{safearcross}. In the second, they can capture explicit requests or confirmations from the \ac{vru}. Prior work on wearable \ac{ar} found that such interfaces can reduce cognitive load in multi-vehicle scenarios and that participants may prefer sending a crossing request to waiting for detection \cite{tran2022wearablear}. An outdoor study by Aleva et al.~\cite{aleva2024outdoorar} showed that head-locked cues maximize visibility over vehicle- or world-locked cues, but may require additional confirmation against the real traffic scene. SafeARCross demonstrated \ac{ar}-based collision warnings and virtual traffic lights using connected vehicles in a real setting; its 20-participant study reported improved perceived safety and reduced cognitive workload compared to an unassisted crossing \cite{safearcross}. Related work on cyclists and pedestrians further shows that \ac{ar} can also support hazard awareness, crossing decisions, and safety-oriented feedback in dynamic road environments \cite{bikear,mrhazardnotifications,towardsfutureinteractions}. Prior work also used a Microsoft HoloLens~2 to reveal occluded connected vehicles~\cite{cooperativear}. The reported delivery latency fell below the 300\,ms timing reference for safety-oriented road applications in ETSI TS 101 539-3~\cite{etsi300ms}.

The literature contains more information on how an \ac{av} should signal to a \ac{vru} than about how a \ac{vru} should initiate and convey an interaction. It also knows more about isolated modalities than about closed, bidirectional systems in which \acp{av} and \ac{vru} can communicate back and forth, expressing intents and receiving acknowledgments. However, these strands are usually evaluated separately. Unlike prior work that evaluates isolated \ac{av}-to-pedestrian signals or isolated input modalities, this paper 
jointly evaluates combined multimodal interaction channels, while examining how people want to communicate with \acp{av}, how the system should make recognition and responsibility visible, and what implementation constraints may interrupt that loop.

\section{Interaction Design}
\label{section:interactions}
The proposed interaction design treats \ac{vru}--\ac{av} communication as a closed interaction loop. Road users normally rely on implicit cues, including vehicle kinematics and pedestrian movement, while occasionally using explicit cues when acknowledgment or negotiation is needed \cite{Lee2021}. % each other's existence by gazing at each other, stepping forward, moving their hands or head, flashing lights, among others. However, when 
Removing the driver leaves vehicle motion intact, but removes person-specific acknowledgment and the ability to signal back. The proposed loop supplements those implicit cues with reciprocal input and feedback.

\subsection{Interaction Goals}
The design goal is %, therefore, to support 
mutual intelligibility: the \ac{vru} should understand the vehicle's current state and intended behavior, while the vehicle-side system should receive the \ac{vru}'s expressed intent. This restores limited human agency, while the \ac{av} retains authority and responsibility for safe navigation. This framing follows prior work, showing that \ac{av}--pedestrian interaction should not be one-directional \cite{colley2021feedbackav}. Building on this view, we define a bidirectional interaction as three actions. First, the vehicle communicates state or intended motion. Second, the \ac{vru} expresses a high-level intent, such as asking the vehicle to stop, proceed, slow down, wait for additional pedestrians, or avoid an obstacle. Third, the system reports what it recognized and whether the \ac{av} accepted or rejected the request.

The three modalities address different parts of this loop. Wearable \ac{ar} is the primary \ac{av}-to-\ac{vru} channel, while vehicle kinematics remain the implicit cue against which its messages should be checked. Unlike a vehicle-mounted display, wearable \ac{ar} can show egocentric, user-specific, and spatially situated information in the \ac{vru}'s own \ac{fov}, which is useful when several road users or vehicles are present \cite{tran2022wearablear,aleva2024outdoorar,safearcross}. Its role is to clarify vehicle state, warnings, and acknowledgments.

Voice interaction is used as an explicit \ac{vru}-to-\ac{av} intent channel. Speech allows the \ac{vru} to express short, high-level requests; therefore, candidate vocabulary focuses on negotiation acts that are plausible in crossing encounters, including stopping, proceeding, slowing down, waiting for pedestrians, and obstacle avoidance. These commands are intentionally framed as semantic intents rather than trajectory instructions, because external road users should not be responsible for detailed vehicle path planning. Additionally, these commands must be interpreted and validated by the \ac{av} before any vehicle action is taken.

Gestures provide a complementary \ac{vru}-to-\ac{av} input channel. They are particularly relevant when the environment is noisy, when speaking is socially inappropriate, when the user cannot or does not want to speak, or when a physical gesture can reinforce the intended meaning of a spoken command. However, gestures are also culturally situated and can be ambiguous when multiple vehicles or pedestrians are present \cite{culturaldronegestures,tran2024gesturemapping}. For this reason, our gesture design emphasizes simple and conventional safety-relevant signals, while treating gestures primarily as an optional channel or disambiguating evidence rather than mandatory or independent commands. Nevertheless, we included multiple gesture candidates for the same interaction intent and gestures that do not pair directly with any voice command. This enabled us to compare alternative gesture mappings for the same intent.

Taken together, the loop proceeds as follows: the vehicle communicates state or intent to a \ac{vru}; the \ac{vru} makes a request through voice, gesture, or both; the system recognizes and transmits the intent directly or via intelligent infrastructure; the \ac{av} validates whether the request is safe and feasible; and \ac{ar} feedback confirms the vehicle response. Vehicle motion then provides an observable cue that corroborates the message. This leads to five design requirements: usable response times; clear and learnable semantics; minimal ambiguity about meaning and target; robustness to noise and occlusion; and visible handling of uncertainty, rejection, or inability to comply.

\subsection{Multimodal Interactions}
\begin{figure*}
    \centering
    \captionsetup[subfloat]{font=scriptsize}
    
    % First row
    \subfloat[Yielding to Cross (I1)]{%
        \includegraphics[width=0.32\linewidth]{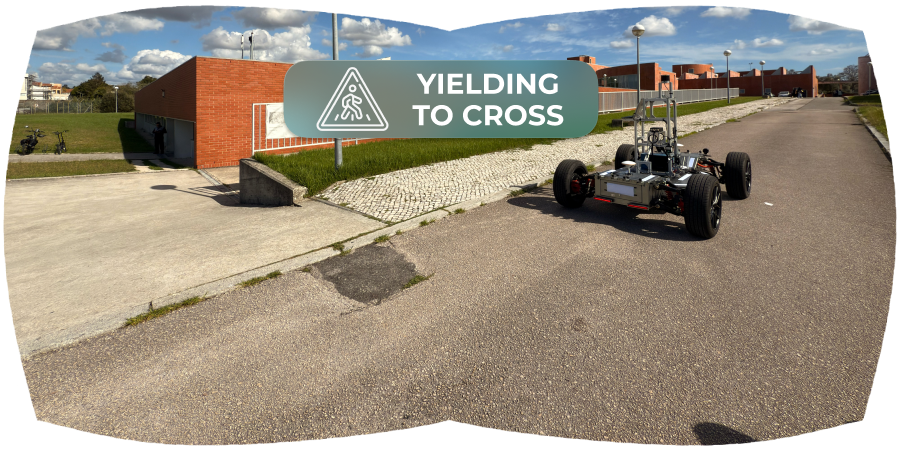}
        \label{subfig:interface1}
    }
    \hfil
    \subfloat[Vehicle Approaching (I2)]{%
        \includegraphics[width=0.32\linewidth]{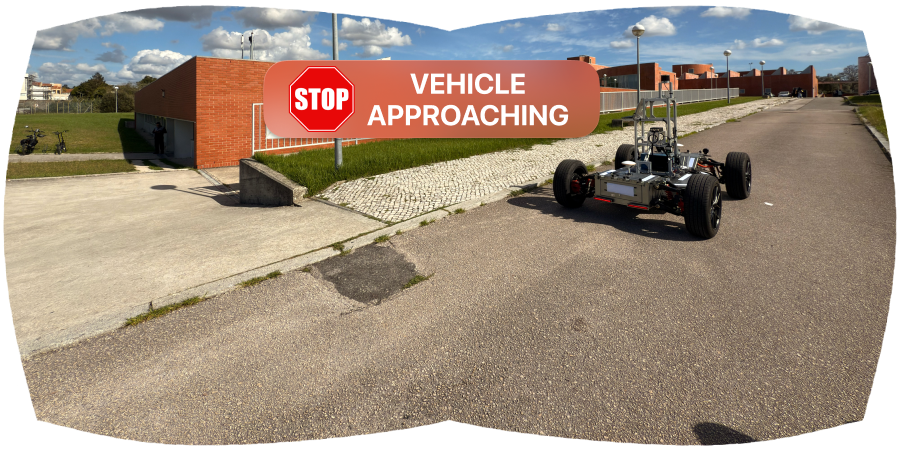}
        \label{subfig:interface2}
    }
    \hfil
    \subfloat[Directional Arrow (I3)]{%
        \includegraphics[width=0.32\linewidth]{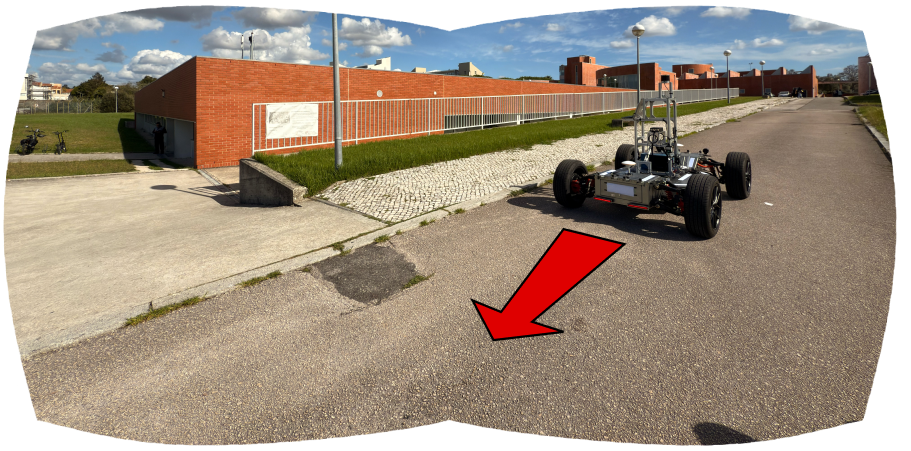}
        \label{subfig:interface3}
    }

    % Second row
    \subfloat[Green Traffic Light (I4)]{%
        \includegraphics[width=0.31\linewidth]{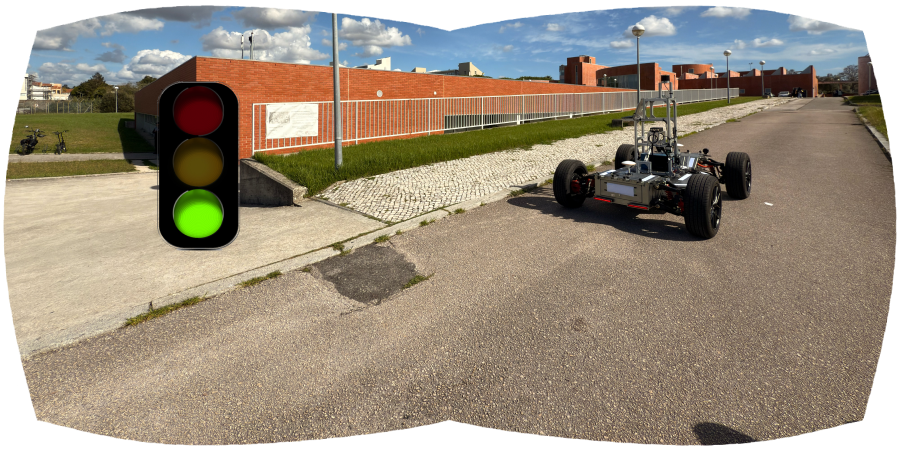}
        \label{subfig:interface4}
    }
    \hfil
    \subfloat[Red Traffic Light (I5)]{%
        \includegraphics[width=0.31\linewidth]{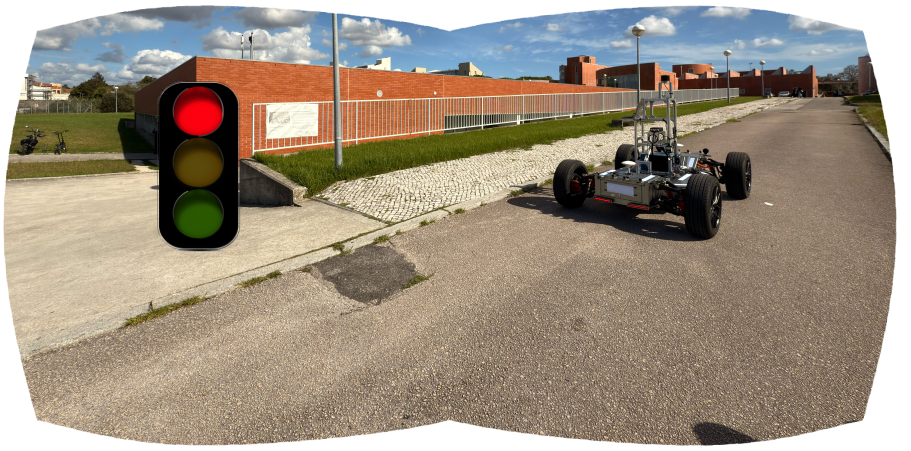}
        \label{subfig:interface5}
    }
    \hfil
    \subfloat[Projected Crosswalk (I6)]{%
        \includegraphics[width=0.31\linewidth]{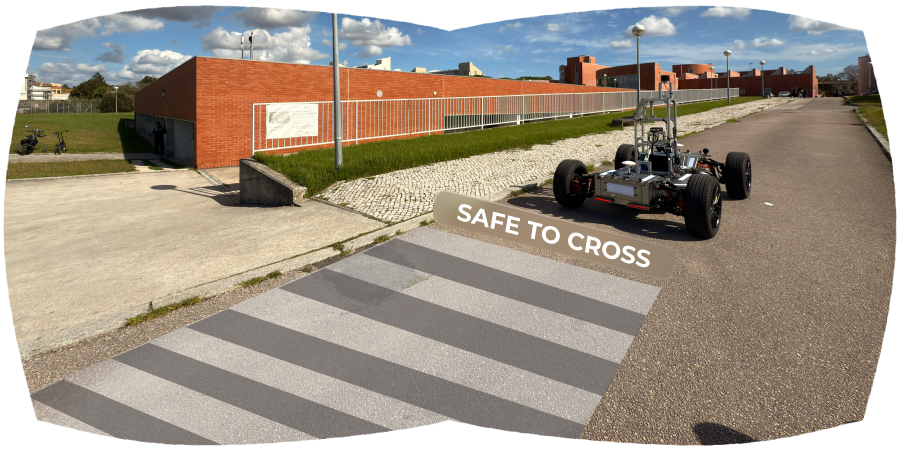}
        \label{subfig:interface6}
    }

    \caption{\label{fig:ar-interfaces}Candidate \acs{ar} interfaces presented in the online questionnaire.}
    \Description{Candidate AR interfaces presented in the online questionnaire.}
\end{figure*}

Six candidates were selected for \ac{av}--\ac{vru} feedback interfaces to cover both directions of interaction. \autoref{fig:ar-interfaces} illustrates them. These are designed to assist pedestrians at crossings using visual notifications, conventional road symbolism, and spatial overlays.

Derived from concept 8~\cite{towardsfutureinteractions}, Yielding to Cross (I1) displayed a textual yielding cue ``YIELDING TO CROSS'' in blue tones, featuring a pedestrian crossing a crosswalk symbol on the left. Likewise, Vehicle Approaching (I2) presented a red warning overlay and a textual alert ``VEHICLE APPROACHING'' with a stop sign on the left side. The stop sign was chosen due to its widespread use; the crosswalk sign, on the other hand, is a bit more ambiguous, but still widely recognizable in its ability to be interpreted as a pedestrian-related task/zone. Forty European countries are part of this convention, making it a recognizable candidate in many European contexts \cite{unitednations}. The \ac{hud} is designed in a notification style in the top center of the user's \ac{fov}, following guidelines for text positions in complex environments, preserving visibility of the walking area \cite{textposition}. The white text contrasts with the colored background plate, offering readability \cite{optimalfontsize, effectsoftext}.

%Based on concept 9, 
Interfaces~I4 and I5 adapt the virtual traffic light cue family implemented in Clérigo et al.'s work~\cite{safearcross}. A green or red 3D traffic light model that appears at the user's leftmost side \ac{fov} to communicate advice through familiar traffic conventions. The traffic light did not include a pedestrian symbol, as it was envisioned to be used also for connected cyclists.

Directional Arrow (I3) was inspired by concept 4, which is meant to represent a field of safe travel~\cite{gibsonfieldofsafetravel}. The field at any given moment represents the paths that a car may take unimpeded, and it would look like a ``sort of tongue protruding forward along the road''~\cite{Kolekar2020}. While Tabone et al.'s approach presents a highlighted road region colored red or green to indicate whether the denoted zones could be safely crossed \cite{dontpanic}, our designs projected an arrow onto the road scene to visualize the vehicle's intended direction of motion. This follows prior work that shows the value of explicit spatial intent visualization \cite{intended_path_projection}, giving additional information to users to decide whether to cross or not.

Based on concept 1, Projected Crosswalk (I6) displays a virtual crosswalk-like overlay aligned with the road surface, representing a spatially anchored safe-crossing cue inspired by previous projected and context-aware crossing interfaces. Instead of being anchored to the road, it is instead anchored to the vehicle, just like concept 1. Additionally, it contains a small billboard-like notification attached to the vehicle, just like concept 2, with a text cue ``SAFE TO CROSS'' in gray tones, just like the crosswalk projected pattern.

Some of these concepts have already been validated in real-world scenarios. A previous study has integrated Interfaces~I2, I4, and I5 on an \ac{ar} safety application~\cite{safearcross} (see \autoref{section:rw}).%into the HoloLens~2 headset to help pedestrians better navigate through traffic by collision warnings and virtual traffic lights. %A user study with 20 participants confirmed that \ac{ar} assistance improved perceived safety and reduced cognitive workload significantly compared to traditional crossings, and that both methods achieved excellent usability. 

The voice commands presented to users were the following: Stop (C1), represented by ``Car, stop''; Proceed (C2), represented by ``Car, go ahead''; Avoid obstacle (C3), represented by ``Car, go around that''; Slow down (C4), represented by ``Car, slow down''; and  Wait for Pedestrians (C5), represented by ``Car, wait for more people''. The command examples were generated through an informal elicitation and refinement process in which candidate phrases were collected from students, researchers, and informal contacts, grouped by intended negotiation act, and normalized into a consistent short-command format. This is consistent with user-defined command elicitation approaches for \ac{av} interaction~\cite{detjen2019userdefined,tran2024gesturemapping}. These phrases are intentionally short and imperative, covering basic negotiation acts that are plausible in pedestrian--vehicle encounters. The use of the word ``car'' acts similarly to a lightweight wake word or directed-speech marker, helping to separate target-directed commands from unrelated speech~\cite{wakeword,wang2020wakeword}. 

\begin{figure}[ht]
    \centering
    \captionsetup[subfloat]{font=scriptsize}
    
    % First row
    \subfloat[Avoid Obstacle (G1)]{%
        \includegraphics[width=0.32\linewidth]{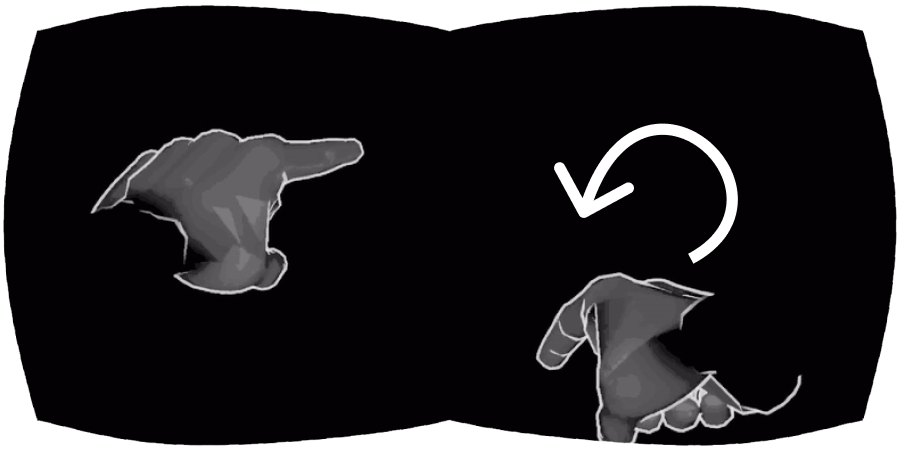}
        \label{subfig:gesture1}
    }
    \hfil
    \subfloat[Emergency Stop (G2)]{%
        \includegraphics[width=0.32\linewidth]{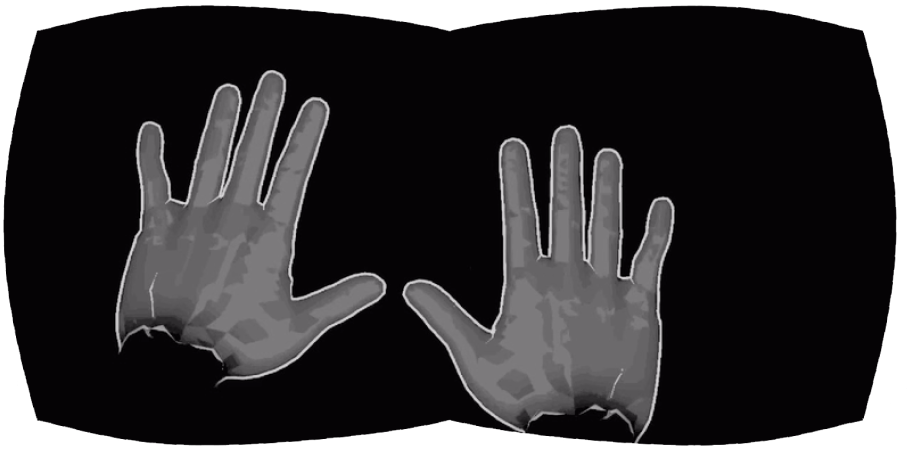}
        \label{subfig:gesture2}
    }
    \hfil
    \subfloat[Follow Me (G3)]{%
        \includegraphics[width=0.32\linewidth]{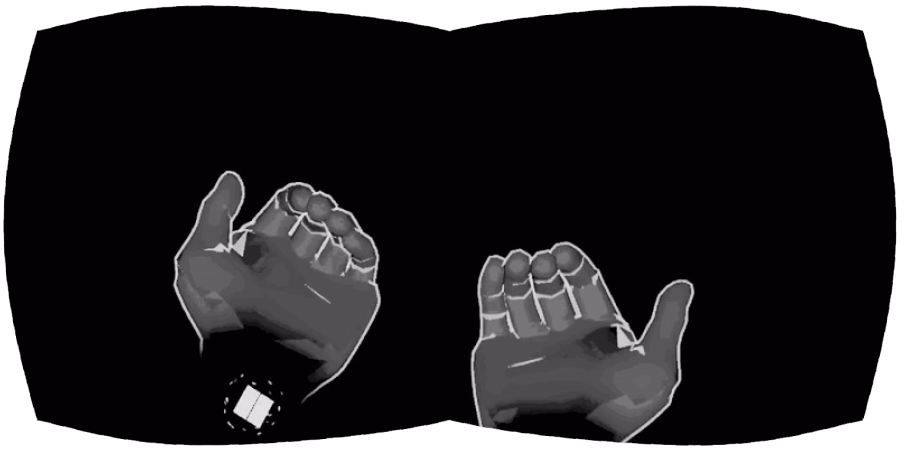}
        \label{subfig:gesture3}
    }

    % Second row
    \subfloat[Go Back (G4)]{%
        \includegraphics[width=0.32\linewidth]{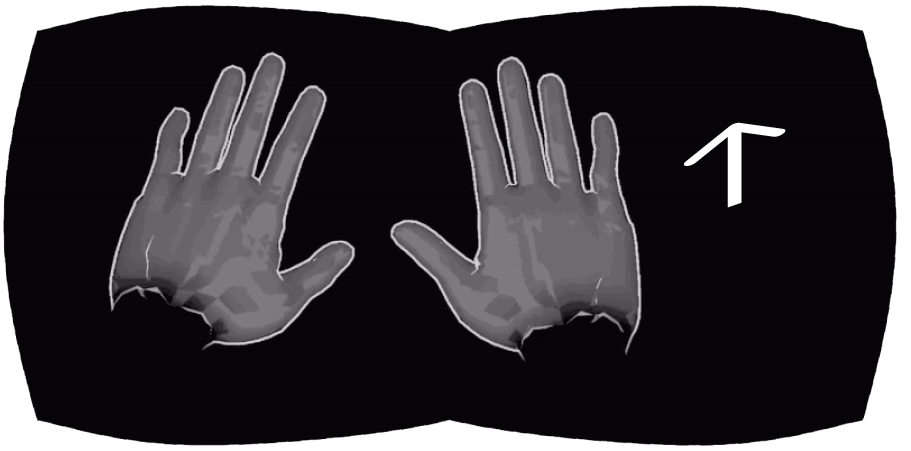}
        \label{subfig:gesture4}
    }
    \hfil
    \subfloat[T-Pose (G5)]{%
        \includegraphics[width=0.32\linewidth]{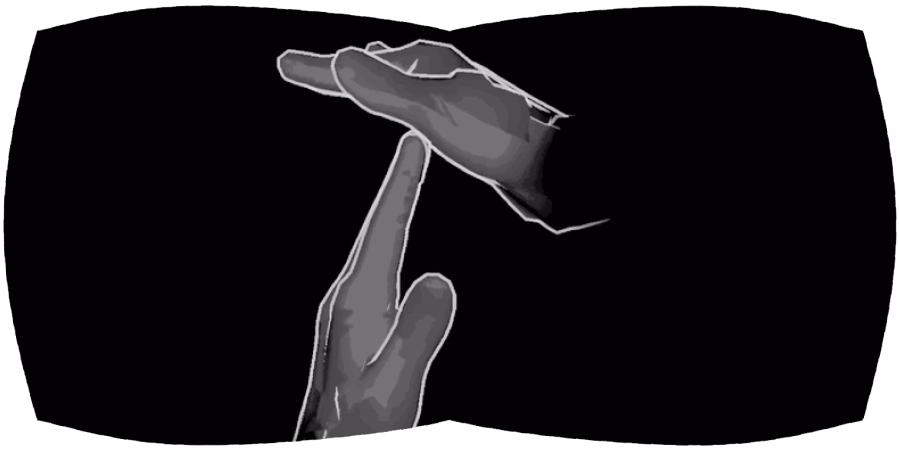}
        \label{subfig:gesture5}
    }
    \hfil
    \subfloat[Proceed (G6)]{%
        \includegraphics[width=0.32\linewidth]{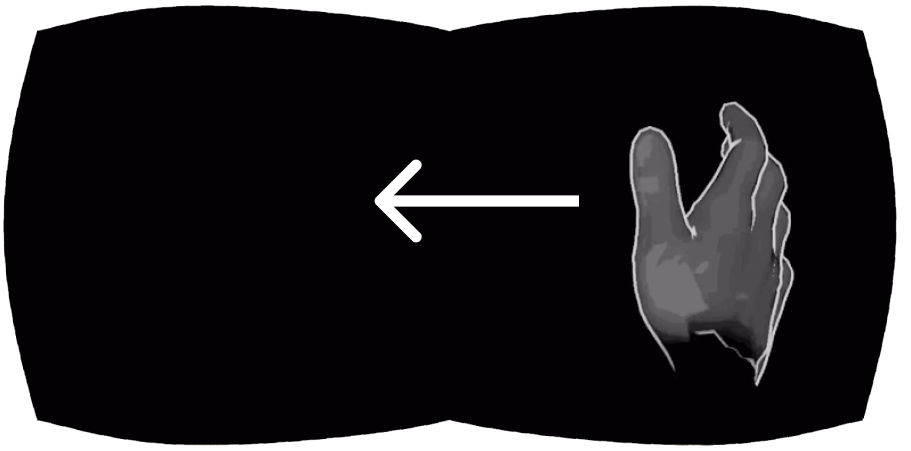}
        \label{subfig:gesture6}
    }

    % Third row
    \subfloat[Slow Down (G7)]{%
        \includegraphics[width=0.32\linewidth]{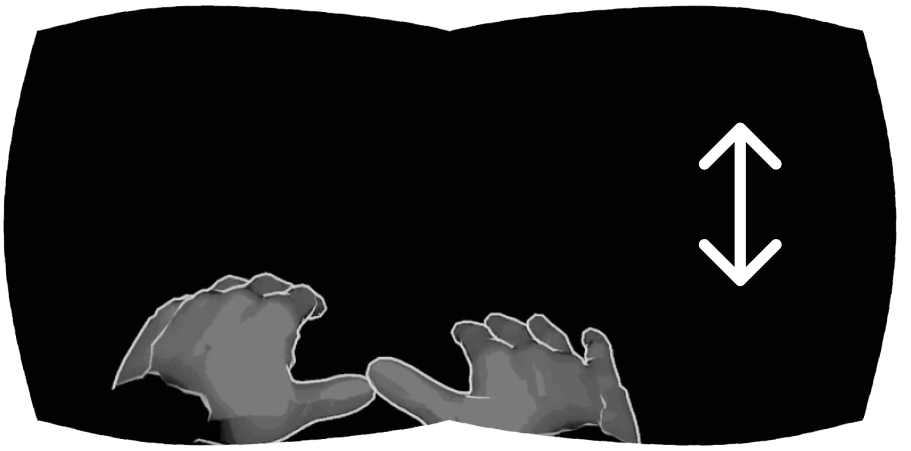}
        \label{subfig:gesture7}
    }
    \hfil
    \subfloat[Stop (G8)]{%
        \includegraphics[width=0.32\linewidth]{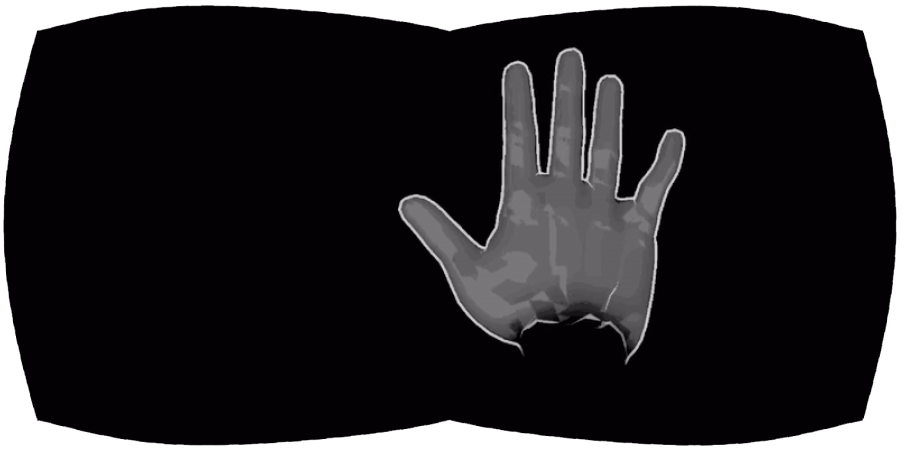}
        \label{subfig:gesture8}
    }
    \hfil
    \subfloat[Wait for Pedestrians (G9)]{%
        \includegraphics[width=0.32\linewidth]{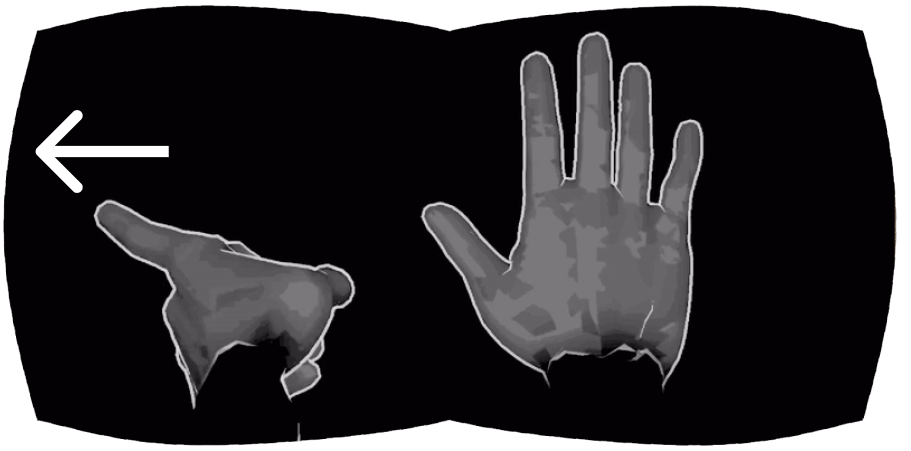}
        \label{subfig:gesture9}
    }

    \caption{\label{fig:gestures}Visual representation of the short videos of gesture alternatives presented in the online questionnaire.}
    \Description{Visual representation of the short videos of gesture alternatives presented in the online questionnaire.}
\end{figure}

The gestures, depicted in \autoref{fig:gestures}, were presented as nine short looping videos for \ac{vru}--\ac{av} commands. These were visual examples of possible \ac{vru} intents that are socially situated and may depend on cultural background, viewing angle, and traffic context~\cite{tran2024gesturemapping,culturaldronegestures}. Avoid Obstacle (G1) used a rotation arm movement intended to indicate that the vehicle should move around an obstruction. Emergency Stop (G2) used two open palms facing the vehicle. Follow Me (G3) used a beckoning or self-directed guiding motion intended to invite the vehicle to move with the user. Go Back (G4) used a backward-directing gesture indicating that the vehicle should retreat or reverse. T-Pose (G5) involved creating a T shape using both palms extended. Proceed (G6) used a forward-directing motion to indicate that the vehicle could continue. Slow Down (G7) used a downward movement associated with ease or calming down. Stop (G8) used a raised palm facing the vehicle, corresponding to a conventional stopping signal. Wait for Pedestrians (G9) used one palm as the Stop (G8) sign, and the other pointing towards the individuals crossing. This set intentionally combines conventional traffic-like movements with more exploratory commands. 

\section{Interaction System Prototype}
\label{section:implementation}
Following the interaction design in \autoref{section:interactions}, the prototype is described to show how human-facing interactions can be connected and where delays or recognition failures can enter the loop. At the interaction level, the system requires the capabilities to capture a person's speech and gestures, recognize and convey the intent to the vehicle, and return an explicit acknowledgment or vehicle response. %the following architectures focus on establishing a wireless communication between two road users: the \ac{vru} equipped with an \ac{ar} headset paired with a computing device, and an \ac{av}. 
We have chosen the Microsoft HoloLens~2 as the \ac{ar} headset, primarily due to its popularity in research and availability.

The communication can be direct between the road users or assisted by cellular or roadside infrastructure, reflecting heterogeneous data transmission between road users within smart cities~\cite{atcll, tahir2021heterogeneous}. The system uses standardized vehicular communication. %Messages sent through \ac{its-g5} follow \ac{etsi}'s standard for \ac{v2x} messages, more specifically, 
Cooperative Awareness Messages (CAMs) \cite{ETSIcam} and \ac{vru} Awareness Messages (VAMs) \cite{ETSIvam} %, and \acp{denm}~\cite{ETSIdenm}. While \acp{cam} and \acp{vam} 
represent road user state, while event-oriented \acp{denm} \cite{ETSIdenm} carry interaction-related requests and responses. %The interaction intent itself is encoded in the payload by tweaking certain message parameters to specific values rather than standardizing a new \ac{denm} cause code.
These standards provide an interoperable implementation enabling different radio technologies and computing locations to preserve the timing and reliability of the interaction.

The implemented interaction loop contains five steps. First, the \ac{av} communicates state or intent through its motion and vehicular messages. % through \ac{v2x} messages, which are received and processed by the \ac{vru}. 
Second, wearable \ac{ar} presents the corresponding information to the person. Third, the person may express a request using speech, gestures, or both. Fourth, the system recognizes, classifies, and transmits the input. Fifth, the vehicle determines whether the request is safe and feasible. Its motion remains the primary observable evidence of execution, while \ac{ar} interface explicitly reports whether the request was accepted or rejected.

\subsection{\acl{ar} Feedback}
\label{subsection:ar-feedback}
\begin{figure}[h]
  \centering
  \includegraphics[width=\linewidth]{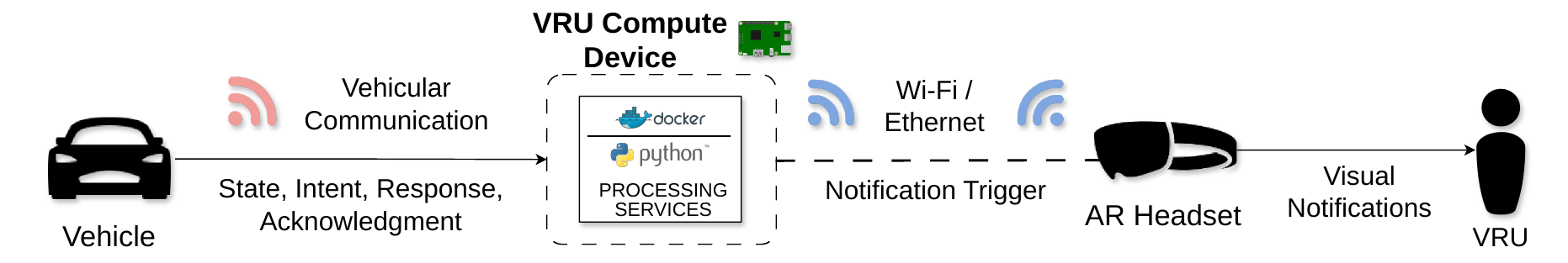}
  \caption{\label{fig:ar-architecture}From vehicle state to human-readable \acs{ar} feedback. The intermediate services implement the path; the user-facing outcome is a notification about vehicle intent, acknowledgment, or response.
  \Description{From vehicle state to human-readable AR feedback. The intermediate services implement the path; the user-facing outcome is a notification about vehicle intent, acknowledgment, or response.}
  }
\end{figure}

\autoref{fig:ar-architecture} depicts the path from vehicle information to personalized feedback. Its main outcome is to place a notification in the user's \ac{fov}; the person should be able to identify which vehicle the information concerns, what it intends to do, and whether a previous request was recognized or accepted. In the prototype, vehicle-side information is sent through vehicular messages and received by a person-carried compute device and forwarded to the headset, which 
%architecture of services and technologies for \acp{vru} to receive an \ac{av}'s intention through \ac{ar}. As shown in the architecture, \ac{v2x} messages emitted by an \ac{av} are received by the \ac{vru} compute device, processed by containerized services, and forwarded to the \ac{ar} headset to trigger a visual notification. The headset 
then displays the corresponding Unity virtual assets approximately 1\,m forward and 10\,cm above the center line of view. 
%These \ac{ar} notifications are designed to capture the user's attention and provide them with real-time, actionable information.

The \ac{e2e} delay imposed by this architecture has already been evaluated in prior work \cite{safearcross,nomsmsba}, and results indicate that it adheres to the standards set by \ac{etsi} \cite{etsi300ms}, mainly the critical 300\,ms \ac{e2e} latency requirement for real-time safety applications. This shows that a warning or acknowledgment arrives before the surrounding traffic situation has changed, supporting an informed decision.

\subsection{Voice Commands}
\label{subsection:voice-interaction}
\begin{figure}[h]
  \centering
  \includegraphics[width=\linewidth]{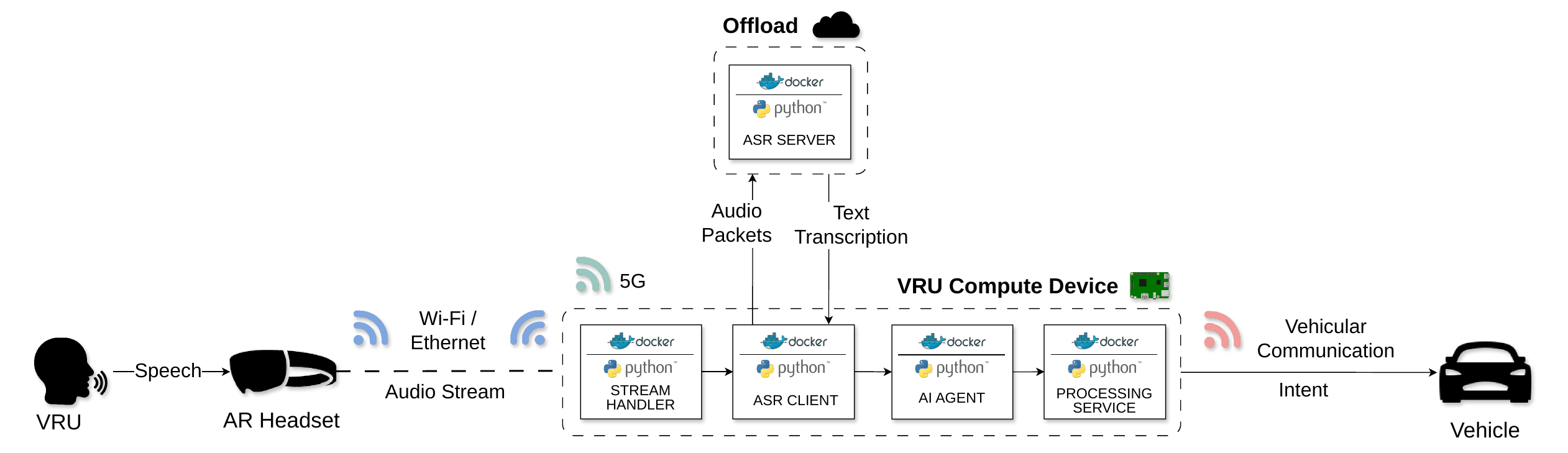}
  \caption{\label{fig:speech-architecture}Speech loop: the person addresses the vehicle, speech is mapped to an intent that is sent for vehicle-side validation. Processing can be local or offloaded.}
  \Description{Speech loop: the person addresses the vehicle, speech is mapped to an intent that is sent for vehicle-side validation. Processing can be local or offloaded.}
\end{figure}

\autoref{fig:speech-architecture} depicts the established architecture for speech interaction. To achieve this, we capture the speech, integrate \ac{asr} to transcribe it, map it to an intent, and convey it to the vehicle for validation \cite{ramos2025shared}.

The interaction's flow starts with a \ac{vru} speaking voice commands while wearing an \ac{ar} headset that captures the audio and relays it to the \ac{vru} compute device. For this purpose, we rely on the HoloLens 2 Sensor Streaming framework %, which enables real-time access to the headset's sensor streams, particularly the microphone~
\cite{silva2025demo, hl2ss}\footnote{\url{https://github.com/jdibenes/hl2ss}}. 
Transcription may then run on the person-carried device or on a nearby remote service. The prototype uses 5G connectivity as a conceptual offloading strategy to place processing wherever the delay and recognition quality is best for user experience.
%The compute device processes this stream and feeds it into an \ac{asr} tool. Using the compute device's 5G connectivity, \ac{stt} is offloaded to a \ac{mec} platform, which contains an \ac{asr} server. This server received audio packets and replies with a live transcription of the \ac{vru}'s speech. 

Finally, the \ac{nlu} unit derives the \ac{vru}'s intention and triggers an interaction event that the processing service translates to the vehicular communication layer. In the current prototype, this event is carried using a \ac{denm}. This design draws from previous HoloLens-based assistive systems that combine perception and audio feedback for object localization~\cite{eckert_object_2018}.

Several tools were tested for \ac{asr}. Vosk~\cite{vosk} offers live transcription, but has limitations in model robustness and accuracy. Moreover, it has less support for multilingual variants than other tools. Whisper \cite{whisper} is OpenAI's state-of-the-art \ac{asr} offering multiple models with different accuracy and latency. However, its original implementation only processes discrete audio, which limits its use in real-time applications. Finally, Kaldi \cite{kaldi}, which is widely used in research, offers a customizable framework. However, for the purposes of this work, its complexity was deemed unnecessary. Considering this, we explored the use of WhisperLive \cite{whisperlive, SileroVAD}, an \ac{asr} tool that combines the accuracy of Whisper and continuous transcription of Vosk. WhisperLive leverages Faster Whisper \cite{fasterwhisper} based on CTranslate2, and integrates voice activity detection through Silero VAD~\cite{SileroVAD}, reducing unnecessary processing during silence and improving Whisper's abilities for real-time recognition. Sherpa-ONNX~\cite{sherpaonnx} is also an alternative for more self-contained speech processing, mainly keyword detections. This makes it a relevant option for future versions that are dependent on embedded systems and do not require the recognition of complex phrases.

For intent recognition, we chose Rasa~\cite{rasa}, which provides the \ac{nlu} layer responsible for mapping natural-language utterances into predefined intents. It was configured with tokenization and lexical features, followed by the Dual Intent and Entity Transformer classifier~\cite{bunk2020diet} with 400 epochs. A fallback classifier has a confidence threshold of 0.8, to prevent uncertain utterances from becoming vehicle-directed messages. This threshold operates on the human-centered safety principle that it is preferable to ask for clarification or to take no action than to make an unintended request.

Given the scenarios described in \autoref{section:online}, we defined Rasa's \ac{nlu} with extensive examples for the five intents: \emph{proceed}, \emph{stop}, \emph{slow down}, \emph{wait for more pedestrians}, and \emph{avoid obstacles}. Each intent contained around 20-30 example phrases, encompassing various natural speech variations. The use of ``car'' in the command examples acts as an addressing cue. This follows the same principle as wake word or keyword-triggered interaction to reduce false activations~\cite{wakeword,wang2020wakeword}. Once an intent is recognized with sufficient confidence, the system publishes the corresponding event to the communication layer, allowing the \ac{vru}'s intention to be encoded and transmitted toward the \ac{av}.

\subsection{Gesture Interaction}
\begin{figure}[h]
  \centering
  \includegraphics[width=\linewidth]{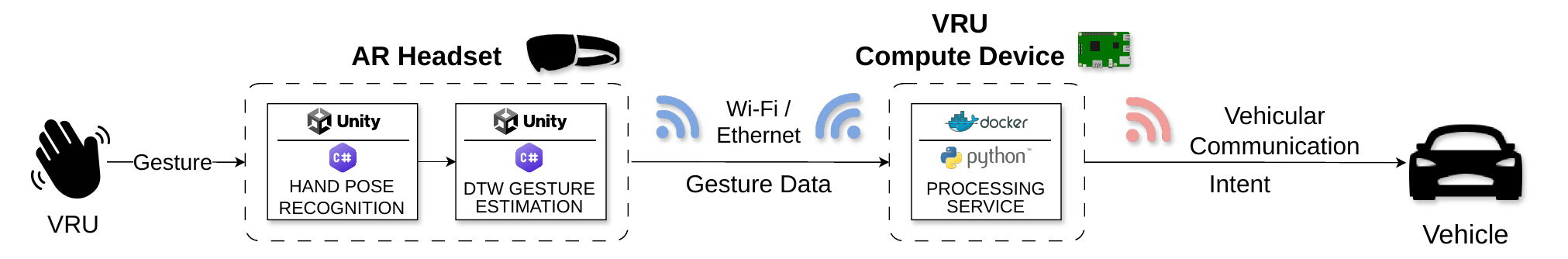}
  \caption{\label{fig:gesture-architecture}Diagram for gesture interaction, with gesture recognition and a vehicular message.}
  \Description{Diagram for gesture interaction, with gesture recognition and a vehicular message.}
\end{figure}

Gesture interaction provides a secondary input channel to complement the aforementioned speech interaction.  As shown in \autoref{fig:gesture-architecture}, the sequence has the person perform the gesture, the \ac{ar} headset estimates whether a known gesture has been performed, and a recognized intent is conveyed for the vehicle via a vehicular message.
%and triggers the corresponding interaction event. The interaction event publishes a message to the \ac{vru} compute device, which publishes the corresponding \ac{v2x} to the \ac{av}.

Hand tracking is performed in Unity/C\# using the built-in hand-tracking capabilities exposed through the mixed reality toolkit 3 (MRTK3)\footnote{\url{https://learn.microsoft.com/en-us/windows/mixed-reality/mrtk-unity/mrtk3-overview}} from Microsoft. Each hand frame stores the 3D positions of the 26 joints across the palm, wrist, and fingers. These frames are accumulated in a sliding buffer while gesture detection is enabled. 

The implementation separates static hand-pose from dynamic gesture recognition. Static gestures are recognized from the instantaneous spatial configuration of the tracked hand joints. This approach is the most suitable for gestures like Stop (G8), Emergency Stop (G2), and T-Pose (G5). 
Dynamic gestures are most suited for gestures with movements such as Avoid Obstacle (G1), Follow Me (G3), Go Back (G4), Proceed (G6), Slow Down (G7), and Wait for Pedestrians (G9). Before attempting classification, the system checks whether the hand has moved sufficiently to reduce false positives from stationary hand poses, and the gesture match occurs from short temporal sequences of hand-joint positions. This distinction follows the broader classification of hand interactions, where static and dynamic poses are treated as different interaction forms with different recognition requirements~\cite{nguyen2023handinteraction}. 

\ac{dtw} is used for gesture matching. The method computes a nonlinear temporal alignment between the live sequence and each stored template, allowing users to perform the same mid-air gesture at different speeds~\cite{sakoe1978dtw}. Previous literature has also used \ac{dtw} for skeleton-based gesture recognition using joint positions \cite{celebi2013skeletondtw}.

%The system also includes a workflow for creating a gesture template. 
Multiple templates can be associated with a target gesture, with each template containing a normalized frame sequence and a configurable \ac{dtw} threshold. This allows for new gestures to be created and then tuned empirically. %The implementation applies a yaw normalization based on the initial palm displacement, aligning the early movement direction to a local reference axis. This preprocessing makes the comparison less dependent on absolute hand position and user orientation. 
At runtime, the recognizer compares the most recent live motion window against each stored template. After a gesture is recognized, the window buffer is cleared, so that the same movement does not get repeatedly triggered. This design supports real-time interaction with high recognition behavior, which is important because ambiguous or incorrectly recognized gestures could otherwise lead to unsafe or confusing \ac{vru}--\ac{av} communication~\cite{gruenefeld2019vroad,chang2024gesture}.

\section{Online Survey}
\label{section:online}
The following section discusses the online survey conducted, outlining participant demographics, methodology, and data collected to evaluate user preference on multimodal interactions with \acp{av}. We structure the evaluation around four research questions:

\begin{description}
    \item RQ1. Which \ac{ar} interfaces are perceived as clearest for conveying an \ac{av}'s intent to a \ac{vru}?
    \item RQ2. Which voice commands and gestures are perceived as most applicable for \acp{vru} to communicate intent to \acp{av}?
    \item RQ3. When voice and gesture are combined, do participants naturally select semantically corresponding gestures for safety-relevant voice commands?
    \item RQ4. What do these preferences imply for the design of a bidirectional multimodal interaction loop between \ac{ar}-equipped \acp{vru} and \acp{av}?
\end{description}

These questions evaluate which of these alternatives were perceived as clearest and most applicable before prioritizing them in the implementation and technical validation. 

\subsection{Participants}
We recruited one hundred and twenty-four participants (N=124), including thirty female and ninety-four male, primarily using social contacts and social media posts. Most participants (76\%) were 20--24 years old. This group is relevant to the likely early adoption of \ac{ar} technologies \cite{qian2020adoption}, and should not be treated as a representative broader population. Because recruitment occurred mainly through Portuguese networks, only seven participants used the English questionnaire; therefore, results should be interpreted as predominantly situated in a Portuguese cultural and linguistic context.

In order to classify the user's affinity with technology, we used the \ac{ati} scale \cite{ati}. The sample had a medium-high affinity for technology, with an average \ac{ati} score of 4.23 (SD\,=\,0.75), and acceptable internal consistency (Cronbach's $\alpha$\,=\,0.80).

\subsection{Design}
The survey was conducted entirely online using LimeSurvey~\cite{limesurvey}, which contained videos and audio samples. All participants conducted the survey independently.

Participants were asked to imagine themselves as \acp{vru} approaching or negotiating a road crossing with an \ac{av}, where no human driver was available for eye contact or informal negotiation. The survey contained three sections, one for each modality: \ac{ar} interfaces, voice commands, and gestures. The scenario framed the \ac{ar} interfaces as information received by the \ac{vru}, and the voice and gesture alternatives as possible ways for the \ac{vru} to communicate intent back to the vehicle. \autoref{section:interactions} describes the voice commands and illustrates the interfaces and gestures presented to participants. %Participants evaluated each interaction for clarity and applicability. At the end of each section, participants ranked their interactions and provided feedback through open-ended questions, allowing insights regarding user experience and potential unexplored interactions.

%The full questionnaire materials, including the surveys used, images, audio prompts, gesture videos, and English/Portuguese versions, are available in the following public repository\footnote{\clerigo{github}}.

\subsection{Measurements and Analysis}
\label{sec:measures-analysis}
The questionnaire combined quantitative and qualitative measurements. For each interaction family, participants rated individual alternatives using 7-point Likert scales. \ac{ar} interfaces were rated for perceived clarity, while voice commands and gestures were rated for perceived applicability. At the end of each modality block, participants also ranked the alternatives to explicitly extract a preference value~\cite{alwin1985measurement}. In the voice--gesture pairing questions, participants selected which gesture they considered most appropriate to accompany each voice command. Finally, optional open-ended questions collected free-text feedback about each interaction family.

For each interaction family, omnibus differences across alternatives were tested using Friedman tests. Kendall's $W$ is reported as the repeated-measures effect size. Where omnibus tests were significant, pairwise post-hoc comparisons were conducted using Wilcoxon signed-rank tests with Holm correction. Participant rankings were reconstructed into preference scores and analyzed with the same Friedman/Wilcoxon procedure.

Gesture selections in the voice--gesture pairing task were analyzed using chi-square goodness-of-fit tests against a uniform distribution, followed by binomial tests for the expected semantic match against chance ($p=1/9$). To assess consistency between ratings and rankings, we computed participant-level Spearman correlations between Likert ratings and the Borda-style ranking preferences.

Open-ended responses were analyzed as exploratory design feedback. We performed preliminary single-coder open coding to identify recurring concerns and suggestions that could help interpret the quantitative trends.

As an exploratory check, we also examined whether modality scores varied with participant characteristics. Gender and questionnaire-language differences were tested using Mann–Whitney U tests, age-group differences using Kruskal–Wallis tests, and associations with \ac{ati} scores using Spearman correlations. These analyses were treated as exploratory because the sample was unevenly distributed across demographic groups. The rank-based tests were selected because the outcomes were ordinal and the demographic groups were uneven; they do not require normally distributed scores. The analyses were treated as exploratory, and small groups were excluded from age comparisons as reported in \autoref{app:sensitivity-checks}, because the recruitment distribution does not support strong population-level demographic claims.

\section{Results}
\label{section:results}
\subsection{Online Survey Results}
We report the results in three stages. First, we summarize preferences for individual \ac{ar}, voice, and gesture alternatives. Second, we examine whether participants paired voice commands with semantically corresponding gestures. Third, we synthesize the statistical and open-ended findings.

Full omnibus and selected post-hoc statistical details are reported in \autoref{app:omnibus-posthoc}.

\subsubsection{Preferences for Individual Modalities}
\label{sec:individual-modality-results}
\begin{figure}[h]
    \centering
    \includegraphics[width=0.7\linewidth]{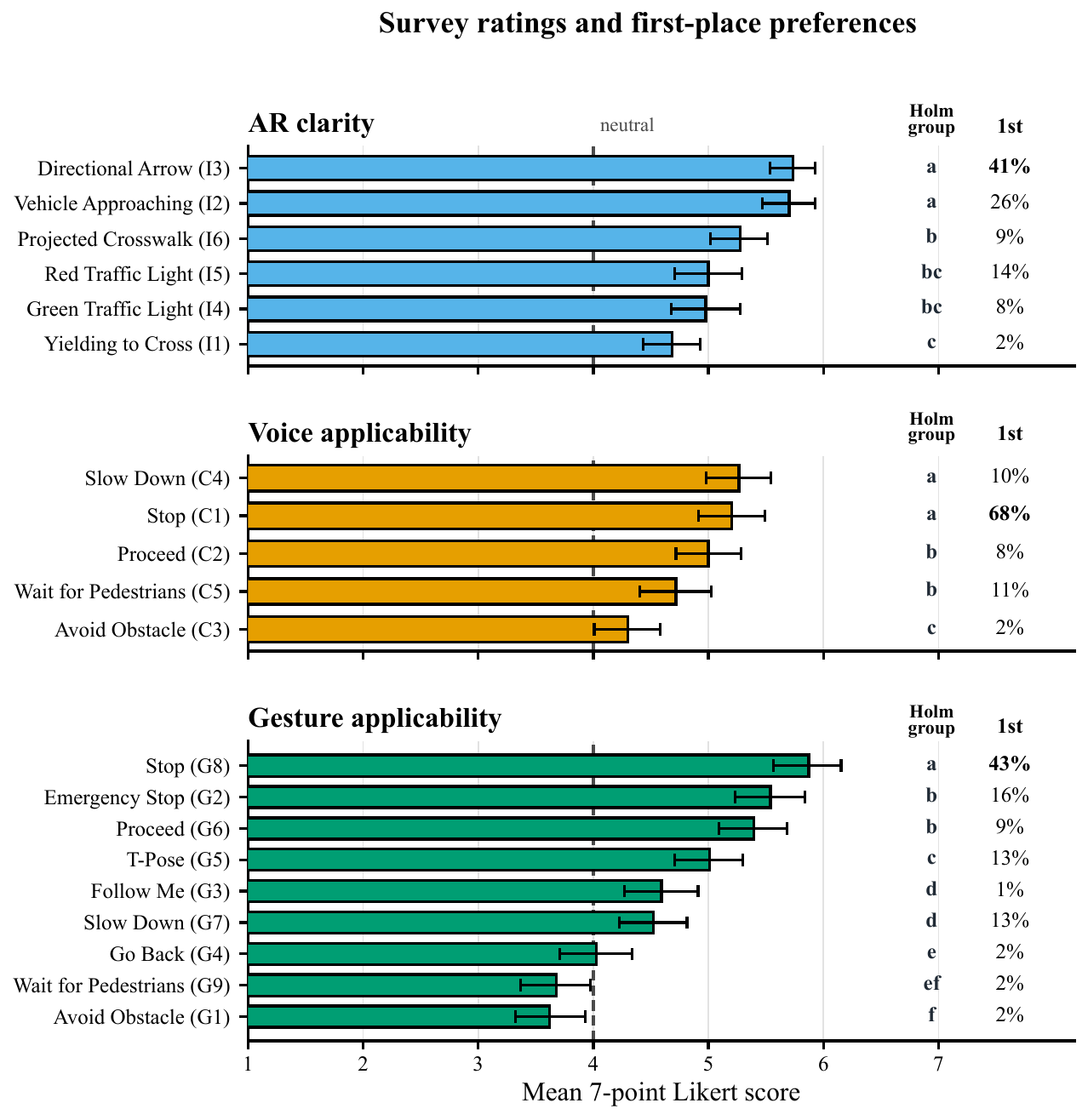}
    \caption{Questionnaire ratings and first-place preferences across interaction families ($N=124$). Ratings use a 1--7 scale, where 7 is the highest perceived clarity or applicability. Bars show mean Likert ratings with 95\% confidence intervals. Percentages on the right indicate how often each item was ranked first. Holm group letters summarize post-hoc Wilcoxon signed-rank comparisons within each family.}
    \Description{Questionnaire ratings and first-place preferences across interaction families ($N=124$). Ratings use a 1--7 scale, where 7 is the highest perceived clarity or applicability. Bars show mean Likert ratings with 95\% confidence intervals. Percentages on the right indicate how often each item was ranked first. Holm group letters summarize post-hoc Wilcoxon signed-rank comparisons within each family.}
    \label{fig:survey-summary}
\end{figure}

The perceived clarity of the \ac{ar} interfaces differed significantly across designs (Friedman $\chi^2(5,N=124)=76.08$, $p<.001$, Kendall's $W=.12$). As shown in \autoref{fig:survey-summary}, the Directional Arrow (I3) ($M=5.73$, median $=6$) and Vehicle Approaching (I2) ($M=5.70$, median $=6$) were the clearest interfaces. They were also the two strongest ranked options, with the Directional Arrow selected as the first option 41\% of times, while the Vehicle Approaching was chosen 25.8\% of times. This indicates that participants generally ranked highly \ac{ar} interfaces that they rated highly. The pairwise post-hoc comparison between these two leading interfaces was not significant after correction. Post-hoc group letters in \autoref{fig:survey-summary} summarize Holm-corrected Wilcoxon signed-rank comparisons within each interaction family; items sharing a letter did not differ significantly.

Revisiting RQ1, results suggest that Directional Arrow (I3) and Vehicle Approaching (I2) are the clearest \ac{ar} interfaces for conveying an \ac{av}'s intention to a \ac{vru}. Nevertheless, all \ac{ar} designs, even the lower-rated \ac{ar} designs, were rated above the neutral midpoint after Holm correction. This pattern suggests that participants understood the general purpose of all visual interfaces, but preferred designs that directly communicated motion direction or imminent warnings.

Voice command applicability also differed significantly across alternatives (Friedman $\chi^2(4,N=124)=97.47$, $p<.001$, Kendall's $W=.20$). The highest mean ratings were observed for Slow Down (C4) ($M=5.27$, median $=6$) and Stop (C1) ($M=5.20$, median $=6$) voice commands, and their pairwise difference was not significant after Holm correction. However, the ranking results revealed a clearer preference, with Stop (C1) ranked first by 67.7\% of participants, while Slow Down (C4) was ranked first by only 10.5\%.

Participants rated several voice commands as applicable, but when forced to choose a preferred command, Stop (C1) emerged as the clear favorite candidate. Avoid Obstacle (C3) was the weakest voice command ($M=4.30$) and did not differ significantly from the neutral midpoint after global Holm correction. This suggests that external users may be uncomfortable issuing commands that are too specific or detailed towards \acp{av}. Voice interaction should therefore emphasize a small set of high-level intents rather than managing commands.

The applicability of the gesture showed the strongest differentiation between the alternatives (Friedman $\chi^2(8,N=124)=398.52$, $p<.001$, Kendall's $W=.41$). The Stop (G8) gesture was the strongest candidate both by rating ($M=5.87$, median $=7$) and by ranking, where it was selected first by 42.7\% of participants and appeared in the top three for 80.6\%. Emergency Stop (G2) ($M=5.54$) and Proceed (G6) ($M=5.40$) were also rated highly. In contrast, Avoid Obstacle (G1) ($M=3.62$), Wait for Pedestrians (G9) ($M=3.68$), and Go Back (G4) ($M=4.02$) were least preferred or near-neutral gestures. Go Back (G4) did not differ from the neutral midpoint, and Avoid Obstacle (G1) showed a small but significant negative deviation from the neutral midpoint after correction.

The gesture results reinforce the importance of using conventional gestures. Overall, well-known road safety metaphors were better rated because they resemble familiar directing movements. By contrast, gestures that require more effort in interpretation and risk ambiguity when used in road traffic are lower rated.

Addressing RQ2, results suggest that the voice commands perceived as the most applicable for \acp{vru} to communicate intent to \acp{av} were Slow Down (C4) and Stop (C1). For gestures, results were much more diverse; however, Stop (G8) emerges as the overall best rated and ranked.

\subsubsection{Voice--Gesture Pairing}
\label{sec:voice-gesture-results}
\begin{figure}
    \centering
    \includegraphics[width=0.7\linewidth]{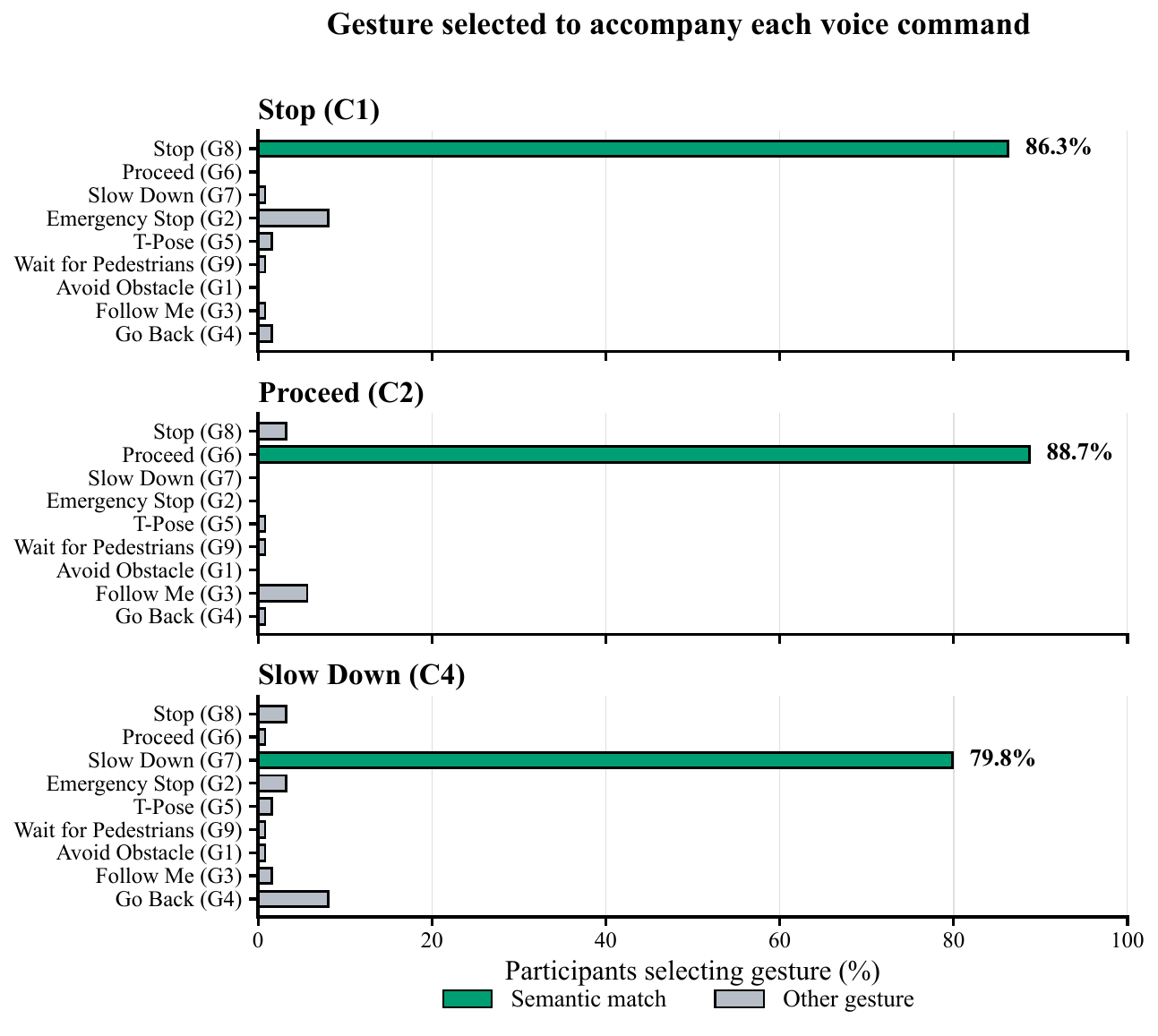}
    \caption{Gesture selected by participants ($N=124$) to accompany each voice command. Green bars mark the expected semantic match.}
    \Description{Gesture selected by participants ($N=124$) to accompany each voice command. Green bars mark the expected semantic match.}
    \label{fig:gesture-matching}
\end{figure}

The voice-gesture pairing task focused on evaluating the best gesture fit to accompany a voice command. \autoref{fig:gesture-matching} shows the selected gestures to accompany three representative voice commands: Stop (C1), Proceed (C2), and Slow Down (C4). These commands were selected because they cover distinct intents. Limiting the task to these representative cases also reduced the participant burden after rating and ranking tasks.

Gesture selections were non-uniform for all three tested voice commands: Stop (C1) ($\chi^2(8,N=124)=715.03$, $p<.001$, Cohen's $w=2.40$), Proceed (C2) ($\chi^2(8,N=124)=759.16$, $p<.001$, $w=2.47$), and Slow Down (C4) ($\chi^2(8,N=124)=597.74$, $p<.001$, $w=2.20$). The direct semantic match was selected by 86.3\% of participants for Stop (C1), 88.7\% for Proceed (C2), and 79.8\% for Slow Down (C4). Exact binomial tests showed that all three direct matches were above chance ($p<.001$). Addressing RQ3, results show that participants showed a strong tendency to pair voice commands with semantically corresponding gestures (\autoref{fig:gesture-matching}).

These results support the multimodal agreement as a safety mechanism rather than redundancy. For example, a spoken Stop accompanied by a raised-palm gesture can provide converged evidence about intent, whereas a spoken mismatched pair should trigger a fail-safe state or no vehicle action.

For each participant, we computed Spearman correlations between item ratings and ranking preference scores within each interaction family. Median participant-level correlations were high for \ac{ar} interfaces ($\rho=.824$), voice commands ($\rho=.718$), and gestures ($\rho=.766$), with all one-sample Wilcoxon tests against zero yielding $p<.001$. This indicates that participants generally ranked highly the same items that they rated highly. Exploratory participant-characteristic checks are summarized in \autoref{app:sensitivity-checks}.

Across the three commands, 316 of 372 selections (84.9\%) corresponded to the expected semantic gesture, while 56 selections (15.1\%) used another gesture. Most of these mismatches were not strongly counter-semantic; therefore, this variation reflects semantically adjacent gestures that could be used for the same command. A small number of counter-semantic pairings were manually inspected; most occurred only once. 

The Slow Down (C4) voice command showed the most variability in gesture pairing, with 25 participants not selecting the expected gesture. However, many of these alternatives can still be interpreted as a slowing down cue: Go Back (G4) was selected by 10 participants, Stop (G8) and Emergency Stop (G2) by 4 participants each, and T-Pose (G5) by 2 participants. All of these imply reducing or preventing forward vehicle motion, but do not express ``slow down'' explicitly. This interpretation could be viewed as participants not requiring the granularity between stopping and slowing down, since stopping-related gestures already imply slowing down. The remaining mismatches were rare: Follow Me (G3) was selected twice, while Proceed (G6), Wait for Pedestrians (G9), and Avoid Obstacle (G1) were each selected once. These mismatches are supported by open-ended comments noting that movement-based gestures could be confused. 

The Stop command (C1) produced 17 non-matching selections, but the majority were still semantically close. Emergency Stop (G2) accounted for 10 of these selections, which is semantically consistent with stopping but uses both hands. T-Pose (G5) and Go Back (G4) were each selected twice, and can also be read as hold-position or retreat signals. Slow Down (G7) was selected once, and it too can be viewed as a way to reflect a halting cue.  
%Participant-level inspection suggests that these isolated cases should be treated cautiously: the Stop--Wait for Pedestrians response came from a participant who missed all three matching tasks and gave non-discriminating gesture ratings.

The Proceed command (C2) produced 14 non-matching selections. The most common was Follow Me (G3), selected by 7 participants, which is plausible because it invites forward movement. The most frequent anomalous case was the pairing with Stop (G8) by 4 participants. One participant explicitly noted in the open-ended questions that the proceed command could be interpreted as signaling a pedestrian to cross, creating a safety risk if the vehicle interprets it; therefore, the users meant to use the gesture for the vehicle to stop so another pedestrian could proceed. The remaining Proceed mismatches, T-Pose (G5), Wait for Pedestrians (G9), and Go Back (G4), each occurred once.

One methodological factor may also have contributed to these mismatches. The matching screen reused numbered labels for gestures, but the display order differed from the earlier gesture-rating section. Therefore, participants who relied on remembered gesture numbers rather than the displayed video could select the unintended gesture. For example, one participant matched Follow Me (G3) with Stop (C1), which corresponded to Gesture 7, which had denoted Stop in the previous gesture-rating section. 

These results point to two sources of gesture ambiguity: (1) semantic overlap between safety-related gestures, and (2) uncertainty about the target of a gesture. Even though most participants paired the expected gesture with the voice command, these findings reinforce the need to treat gestures as confirmation input rather than isolated commands. This agreement can increase confidence, while mismatched signals should make uncertainty visible and trigger confirmation or a fail-safe response.

\subsubsection{Open-ended Feedback}
\label{sec:open-ended-feedback}
In total, 29 participants commented on the \ac{ar} interfaces, 13 commented on voice commands, and 16 commented on gestures.

The open-ended \ac{ar} feedback clarifies why some familiar-looking interfaces did not dominate the ratings. A recurring concern was that traffic light symbols could be misinterpreted. There were 18 participants who explicitly suggested replacing generic traffic lights with pedestrian traffic-light symbols. Eight participants noted that the projected crosswalk could be misread if its orientation did not align with the pedestrian's crossing direction. Six participants suggested combining or layering cues, with one suggesting pairing a directional arrow with an approaching vehicle warning, so that the \ac{vru} sees both the vehicle's intended motion and the required pedestrian action.

As described in \autoref{section:interactions}, the traffic lights' interfaces had unmarked bulbs because the design was intended to remain applicable beyond pedestrians, including connected cyclists and other \acp{vru}. The open-ended responses showed the trade-off of that choice, since the improvement in cross-\ac{vru} compatibility reduced pedestrian-specific clarity in a crossing task. A practical solution would be to make the symbol changes according to the detected or self-declared road-user type. This would preserve the broader \ac{vru} framing without forcing pedestrians to interpret a symbol that they associate primarily with vehicle traffic.

Regarding voice commands, 4 participants stressed that commands could be too short and not acoustically distinct. Detailed commands, such as obstacle avoidance, were criticized as ambiguous or as placing the pedestrian in the role of coordinating traffic, appearing in 4 responses. This supports the quantitative results and indicates that some users may not want the burden nor the capability to coordinate traffic. Broader concerns about voice reliability, misuse, or conflicting commands from multiple \acp{vru} appeared in 6 responses. 

Gesture comments emphasized ambiguity, as 8 participants noted that gestures that share an initial posture or overall momentum could be difficult to remember. Three participants suggested favoring the use of simpler, faster, or one-handed gestures. Together with the quantitative results, these comments indicate that gestures should only be used when they are simple and clear, and that the \ac{ar} system should always confirm whether the intended command was recognized.

No participant explicitly mentioned false activations. Because the questionnaire presented fixed examples rather than a live recognizer, this absence should not be interpreted as evidence that unintended activation is unimportant.

\subsection{Prototype Feasibility Results}
The prototype evaluation asks whether the candidate modalities can work as a closed loop with timing and recognition characteristics that justify a subsequent human evaluation of the situation.
%As mentioned in \autoref{subsection:ar-feedback}, the latency and architecture of \ac{vru}--\ac{av} have already been validated; therefore, this section focuses on testing the latency associated with the \ac{asr} offloading system and its accuracy. %\clerigo{Is this the correct framing? Testing gesture recognition would implicate gather data from different user ans generate something like a confusion matrix. Gesture recognition is included in the system architecture as a prototype input channel, but we do not report a gesture-recognition accuracy benchmark in this paper. The technical validation should therefore be read as evidence for the speech and \ac{v2x}-enabled intent interaction, complemented by an integrated demonstration of the broader loop.}

\subsubsection{Latency and Recognition}
Vehicle-initiated \ac{ar} feedback was evaluated in prior work and was not rebenchmarked here. In a peer-to-peer configuration using the European wireless vehicular communication standard ITS-G5, delivery to the \ac{ar} display required \(81.53 \pm 8.36\)\,ms \ac{e2e}~\cite{nomsmsba,cooperativear}. This prior result falls below the 300\,ms reference for safety-related applications ~\cite{etsi300ms}.

\begin{figure}[ht]
  \centering
  \includegraphics[width=0.8\linewidth]{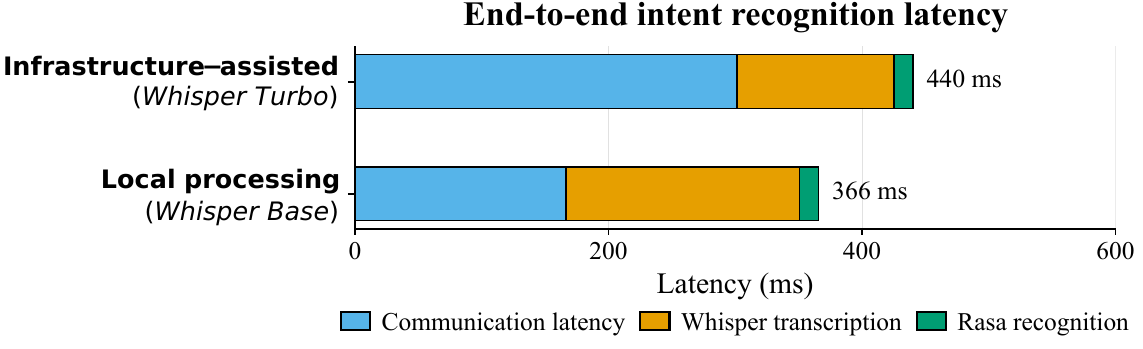}
  \caption{\acs{e2e} intent-recognition latency for infrastructure-assisted processing using Whisper Turbo and local processing using Whisper Base. The measurement ends when the intent is recognized. \autoref{app:interaction-benchmarks} reports the benchmark details.}
  \Description{E2E intent-recognition latency for infrastructure-assisted processing using Whisper Turbo and local processing using Whisper Base. The measurement ends when the intent is recognized. Appendix B reports the benchmark details.}
  \label{fig:e2e-latency}
\end{figure}

In the controlled setup, retrieving the HoloLens microphone stream introduced approximately 166\,ms under both wired and Wi-Fi conditions. The infrastructure-assisted processed speech pipeline reached 440.04\,ms. The representative local configuration reached an \ac{e2e} intent-recognition delay of 365.63\,ms.  \autoref{fig:e2e-latency} decomposes these delays for two representative configurations: infrastructure-assisted \ac{asr} using the Turbo model from Whisper and local \ac{asr} using the Base model from Whisper. These configurations differ in both computing placement and model.

Recognition quality was assessed using \ac{wer} and \ac{cer}. On controlled command audio, a large model on the infrastructure-assisted configuration achieved 2.9\% \ac{wer} and 1.6\% \ac{cer}, compared with 7.0\% and 5.5\% for a smaller local model. With added urban noise, these values became 3.4\% and 2.0\% versus 8.0\% and 6.7\%, respectively. The best configuration detected the specified wake word in 96.77\% of the commands with no false positives in that controlled dataset. Full hardware specifications, distributions, model results, and evaluation procedures are retained in \autoref{app:interaction-benchmarks}.

The \ac{dtw} gesture templates contain varying numbers of frames with 11 joints per frame. At the configured 20\,ms fixed-update interval, their required motion-sampling time depends on the gesture. Once the final sample is available, matching and event dispatch occur on that update, with at most 20\,ms of scheduling delay.

The communication delay has different meanings for the two representative configurations. For local processing, it corresponds to the microphone latency. For \ac{mec} offloading, it includes 5G transport and the overhead introduced by adapting the service to support offloading. Communication delay also encompasses all inter-service transport latency, \acs{its-g5} broadcast, and vehicle-side command interpretation; vehicle actuation has already been explored in previous work \cite{ramos2025shared,nomsmsba}. Rasa intent recognition remained approximately constant, around 15\,ms.

The findings do not identify one preferred computing placement. In the representative configurations presented, local processing produced lower \ac{e2e} latency but higher transcription error, whereas the infrastructure-assisted configuration reduced transcription error with an additional communication overhead. From a human-centered perspective, this shifts risk between waiting longer without feedback and being misunderstood more often. An infrastructure-neutral system should adapt when possible and, regardless of placement, use confidence gating, agreement across modalities when available, and explicit \ac{ar} feedback that reports what was recognized and whether the vehicle accepted or rejected the request. Low confidence or conflicting inputs should result in clarification or no vehicle-side action.

\subsubsection{Integrated Demonstration} A demonstration was conducted on the \ac{atcll} 
%[anonymized living lab] 
platform~\cite{atcll} 
in Portugal. The demonstration video%[removed for anonymity — provided in Supplementary Material]
\footnote{\url{https://youtu.be/cYHqxrlKcZg}} 
shows a \ac{vru} wearing a Microsoft HoloLens 2 and a Raspberry Pi 5 compute device interacting with a PIXKIT~2.0\footnote{\url{https://www.pixmoving.com/pixkit}} Autonomous Driving Development Kit. Speech, gesture, vehicular communication, vehicle-side handling, and \ac{ar} acknowledgment completed the implemented loop outdoors. This closes an important feasibility gap left by survey-only and simulation-only work.

\section{Discussion and Limitations}
\label{section:discussion}
The survey and prototype provide complementary evidence about candidate interactions and their implementability. This work extends evidence that pedestrians value feedback after initiating communication~\cite{colley2021feedbackav} by adding an attributable wearable \ac{ar} and a multimodal path to a physical autonomous platform. Representative systems addressed complementary parts of the loop: \emph{VoGe}~\cite{saurasperez2017voge} combined speech and pointing for vehicle occupants; \emph{I See Your Gesture}~\cite{epke2021gesture} paired pedestrian gestures with a vehicle response in \ac{vr}; and \emph{SafeARCross}~\cite{safearcross} provided wearable \ac{ar} feedback without a pedestrian-to-vehicle input. This present prototype combines external speech and gesture requests with wearable feedback and vehicle-side validation in one outdoor demonstration.

\subsection{Design Implications}
\label{sec:design-implications}
Addressing RQ4, the combined questionnaire and prototype results lead to the following design implications.

\paragraph{Prioritize explicit \acs{av} motion intent in \acs{ar}.}
The strongest \ac{ar} interfaces represented future vehicle behavior through a projected Directional Arrow (I3) and a Vehicle Approaching (I2) warning. Visual feedback should therefore avoid symbolic cues when a clear motion cue or explicit warning can be shown.

\paragraph{Adapt traffic light symbols to the road-user role.}
Eighteen respondents requested pedestrian-specific traffic light symbols. At pedestrian crossings, pedestrian iconography should be used rather than an unmarked traffic light. A broader \ac{vru} system should adapt the symbol to a detected or self-declared road user role.

\paragraph{Keep voice commands at the intent level.}
Participants favored Stop (C1), Slow Down (C4), and Proceed (C2) over more detailed commands. Voice should express a negotiation intent rather than prescribe a vehicle trajectory. 

\paragraph{Keep Gestures simple and conventional.}
Stop (G8), Emergency Stop (G2), and T-Pose (G5) were rated higher because they resemble familiar road traffic movements. These are also simpler gestures that require less motion than the other alternatives shown to the user. Gesture vocabularies should therefore remain small, safety-relevant, and easy to perform and recognize.

\paragraph{Treat gestures as disambiguation signals.}
Participants usually paired speech with semantically corresponding gestures. The technical system can exploit this by increasing confidence when modalities agree and entering a fail-safe when they conflict. Gestures should remain optional because traffic noise, occlusion, lighting, and social context affect which modality is viable.

\paragraph{Make recognition, uncertainty, and responsibility visible.}
The interface should distinguish at least four states: input received, intent recognized, request accepted or rejected, and vehicle response in progress. This prevents silence misinterpretation. The \ac{av}, not the \ac{vru}, remains responsible for deciding whether and how to act.

\paragraph{Explicit feedback must remain consistent with vehicle kinematics}
Because the timings of messages can alter crossing decisions and create overtrust~\cite{declercq2019ehmi,hollander2019overtrust}, explicit feedback should progress through the interaction stages and match the vehicle's real-life actions.

\begin{table*}[h]
\centering
\small
\caption{Literature-grounded timing and recognition reference points, prototype results, and their human-facing interpretation.}
\label{tab:interaction-requirements}

\begin{tabularx}{\textwidth}{
    >{\raggedright\arraybackslash}p{0.11\textwidth}
    >{\raggedright\arraybackslash}X
    >{\raggedright\arraybackslash}p{0.25\textwidth}
    >{\raggedright\arraybackslash}p{0.3\textwidth}
}
\toprule
Interaction & Reference & Prototype evidence & Human-facing implication \\
\midrule

Speech\newline acknowledgment
&
Conversational turns normally have gaps near 200\,ms~\cite{levinson2015turntaking}; delays of 0--250\,ms were rated most favorably in one automotive conversational-agent study~\cite{usablevoicedelay}.
&
Speech intent recognition took 365.63\,ms with local computing and 440.04\,ms with infrastructure assistance.
&
Recognition is slower than a normal conversational response, so the interface should acknowledge audio capture akin to voice assistants~\cite{amazonAlexaAutomotive}.
\\

\addlinespace

%Context-dependent\newline voice response
%&
%An expert simulator study found that latency preferences depend on task context; 1.5\,s was rated most favorably among the tested values, and responses up to 5\,s remained acceptable for satisfaction~\cite{sogemeier2024latency}.
%&
%All interaction paths complete below 0.5\,s before vehicle decision and return feedback.
%&
%The interaction is within the range that can remain usable.
%\\
%\addlinespace

Vehicle-initiated\newline \ac{ar} feedback
&
ETSI states that the total \ac{e2e} latency for a collision-risk warning is 300\,ms~\cite{etsi300ms}.
&
Prior peer-to-peer ITS-G5 delivery to an \ac{ar} display required \(81.53 \pm 8.36\)\,ms \ac{e2e}~\cite{nomsmsba,cooperativear}.
&
When not requiring user input recognition, latency is comfortably met. The cue must still be noticed, understood, and consistent with vehicle motion.
\\

\addlinespace

User-initiated\newline request
&
The 300\,ms value concerns warning freshness, not a universal threshold for every interaction stage~\cite{etsi300ms}.
&
Speech intent required 365.63--440.04\,ms. Gesture matching depends on the template, it takes at most 20\,ms after the final sample.
&
Neither input is an emergency control or collision warning; vehicle perception, prediction, and braking remain independent.
\\

\addlinespace

Command\newline recognition
&
A multimodal in-vehicle command recognizer combining voice and radar reported 96.4\% command accuracy under noisy conditions~\cite{oh2021commandfusion}. 
&
Best controlled command transcription: 2.9\% \ac{wer}, 1.6\% \ac{cer}, and 7.0\% \ac{wer} and 5.5\% \ac{cer} when using local processing.
&
Word-level errors are not intent-level errors. Although results suggest that recognition is within range. Further in-field studies may help understand speech and language variability.
\\

\bottomrule
\end{tabularx}
\end{table*}

\subsection{Human-facing Interpretation of Prototype Performance}
There is no single latency requirement that applies to every stage of the proposed interaction. Human conversation, an automotive voice assistant, a negotiated road request, and a collision warning operate under different expectations and consequences. Likewise, the literature does not establish a universal \ac{wer} or \ac{cer} threshold at which an external road user command becomes safe. \autoref{tab:interaction-requirements} presents reference points for these requirements, and compares them to the prototype's results.

The comparison yields two additional design implications. First, the system should expose an immediate ``input received'' state, because speech intent recognition arrives later than the timing of an ordinary conversational agent. Second, the speech interaction must remain a request-and-acknowledgment channel. It cannot replace the vehicle's independent perception and collision-prediction functions, because it exceeds the 300\,ms threshold~\cite{etsi300ms}.

\subsection{Limitations}
\label{section:limitations}
This work has several limitations that should guide the interpretation of the results. First, the online survey evaluated perceived clarity and applicability rather than situated crossing behavior. Participants %judged images, videos, and command examples while imagining an \ac{av} encounter; they 
did not experience risk, time pressure, vehicle motion, road noise, social presence, or a real street crossing task. The results therefore identify promising candidates and principles but do not demonstrate effects on behavior, trust, safety, or workload%that these candidates improve crossing decisions, response time, trust in automation, perceived safety, or perceived workload in traffic.

Second, the participant sample was
%not demographically balanced and pedestrian-oriented. The sample 
was young, mostly Portuguese, male-skewed, and pedestrian-oriented. %This age distribution may partly reflect greater familiarity with emerging technologies and the likely profile of early adopters of wearable \ac{ar}, but it limits generalizability to cyclists, older \acp{vru}, children, and people with disabilities. 
The exploratory age effects reported in \autoref{app:sensitivity-checks} further suggest that demographic differences should not be treated as negligible. 
%especially because road-crossing ability, risk perception, visual acuity, hearing, and comfort with speech or gesture interfaces can vary substantially across the population.

Third, the interaction vocabulary is culturally and contextually situated. Apart from the speech commands being language-sensitive, some signs, traffic light conventions, and gestures may be familiar in Portugal or broader European contexts, but may not transfer directly to other regions or road cultures. %such as Asia. This limitation is especially important for gestures because, apparently, simple movements, such as waving, pointing, or raising a palm, can differ in politeness, urgency, or meaning across cultures and traffic norms.

Fourth, the speech-recognition evaluation used scripted, \ac{tts}-generated command audio and 
%for the command-oriented \ac{nlu} condition. This allowed controlled comparisons across \ac{asr} models and computing configurations, but it 
it did not capture spontaneous speech, accents, hesitations, overlapping speakers, or wind. %The \ac{fleurs} condition provides a broader unconstrained-speech benchmark, but it does not follow the system's command grammar or wake word structure. 
The reported \ac{wer}, \ac{cer}, and wake word metrics in \autoref{app:interaction-benchmarks} should therefore be interpreted as controlled pipeline indicators rather than as field performance estimates.

Fifth, the current work evaluates one person interacting with one focal vehicle. Real streets contain multiple road users whose requests may overlap or conflict; therefore, a scalable deployment must determine who is addressing whom and prevent intents being attributed to the wrong road user.

Finally, the current hardware setup may influence both usability and measured performance. A HoloLens~2 paired with an external compute device is suitable for research prototyping, but it is heavier and more intrusive than a future consumer-scale system. The hardware burden may affect walking comfort, field of view, social acceptability, and willingness to use gestures or speech near traffic. Since most \ac{ar} interfaces do not rely on 3D mapping, a consumer-ready alternative could leverage smart glasses with a \ac{hud}, paired with a smartphone with cellular connectivity. 

\section{Conclusions and Future Work}
\label{section:conc}
This paper investigated how \ac{ar}-equipped \acp{vru} can participate in bidirectional multimodal interactions between \acp{vru} and \acp{av}, rather than relying only on vehicle-side cues. We combined candidate design synthesis, an online survey with 124 participants, a prototype spanning wearable \ac{ar}, speech and gesture recognition, and vehicular communication, controlled technical benchmarks, and an outdoor demonstration. Together, these validate the interaction from human preference through technical feasibility.

The online survey results favored preferred \ac{ar} cues that made vehicle motion or warnings explicit, high-level speech requests such as stopping or slowing down, and simple, conventional gestures. Participants showed a strong tendency to pair semantically corresponding gestures with spoken commands, supporting multimodal redundancy as a design principle. The prototype completed the intended path from \ac{vru} input, through recognition and vehicular communication, to vehicle-side handling and \ac{ar} acknowledgment. Controlled speech benchmarks further exposed a trade-off between the lower latency of local processing and the lower transcription error of infrastructure-assisted processing.

Taken together, these findings support that \ac{vru}--\ac{av} communication should be designed as a closed interaction loop. A \ac{vru} can express an intent, the system reports what it recognized, the vehicle independently determines whether that request can be acted upon safely, and the outcome is communicated back to the \ac{vru}. This framing preserves human agency without transferring responsibility for safe vehicle behavior to the external road user.%, rather than a one-way display solution. 

The paradigm shift is that automated mobility becomes capable of reciprocal, attributable, and inclusive interaction with the people whose safety and freedom of movement depend on it.

Future work should evaluate a complete interaction loop with participants in a controlled outdoor experiment using \ac{ar} and an \ac{av}. Such studies should measure crossing behavior, trust in automation, self-perceived workload, response time, and time-to-cross. They should also distinguish whether feedback was noticed correctly, understood, and believed, examining the consequences of delayed, rejected, or incorrect recognition. Moreover, future work should test broader participant groups, including cyclists, older adults, children, people with impairments, and with different levels of technology familiarity. Beyond cultural variation, future work should examine accessibility, social acceptability of speaking and gesturing in public, and wearable heterogeneity. Additionally, user applications should calibrate these cues when first encountering or registering. Users could be prompted to speak their preferred command variants and perform key gestures several times; this way, the system could be tailored to the users' preferences, improving gesture- and voice-recognition accuracy for outdoor environments. 

Future technical work should move from component feasibility toward robust in-the-wild recognition. The interaction should also be extended beyond the current one-\ac{vru}, one-\ac{av} scenario to investigate target attribution and conflicting requests when multiple road users and vehicles are present.

%\begin{acks}
%\section*{Acknowledgments}
%This work was supported in part by the EU's HE research and innovation programme HORIZON-JU-SNS-2023 under the 6G-PATH project (Grant No. 101139172). We thank all study participants for their time and feedback.
%\end{acks}

\section*{Open Science}
Anonymized data, questionnaires, and evaluation scripts are available 
%\revision{through a persistent repository before publication,} 
in this repository\footnote{\url{https://github.com/nap-it/multimodal-vru-av-interactions}}, in accordance with applicable ethics guidelines.
%Anonymized data, questionnaires, and evaluation scripts are available as supplementary material for reviewers.

\section*{Ethics Statement}
This research involved human participants through an online questionnaire. Under the institutional guidelines applicable to this study, formal ethics committee review or approval was not required. Participation was voluntary.

\begin{comment}
\color{red}
%TC:ignore
\ifshowwordcount
  {\color{red}
    \section*{Draft Word Count}
    Main text word count, estimated using \texttt{texcount}:
    \textbf{\input{main-words.sum} words}
  }
\fi
%TC:endignore
\color{black}
\end{comment}

%TC:ignore
\bibliographystyle{ACM-Reference-Format}
\bibliography{references}

%TC:ignore

\appendix
\section{Supplementary Questionnaire Statistics}
\label{app:questionnaire-statistics}
This appendix reports supplementary statistical details for the questionnaire analyses. The tables below provide the omnibus tests, selected post-hoc contrasts, and exploratory sensitivity checks.

\subsection{Omnibus and Post-hoc Tests}
\label{app:omnibus-posthoc}
\autoref{tab:questionnaire-omnibus} reports the Friedman omnibus tests on ratings and preference scores. \autoref{tab:selected-posthoc-contrasts} reports selected Holm-corrected pairwise contrasts that support the main comparisons discussed in the results. The selected contrasts focus on leading candidates, close competitors, and weakest alternatives.% rather than listing every pairwise comparison.

\begin{table}[!h]
    \centering
    \small
    \caption{Omnibus Friedman tests for questionnaire ratings and reconstructed preference rankings. Higher scores indicate stronger preference. All omnibus tests were significant at $p<.001$.}
    \label{tab:questionnaire-omnibus}
    \begin{tabular}{llrrr}
        \toprule
        Measure & Family & df & $\chi^2$ & Kendall's $W$ \\
        \midrule
        Likert rating & AR clarity & 5 & 76.08 & .12 \\
        Likert rating & Voice applicability & 4 & 97.47 & .20 \\
        Likert rating & Gesture applicability & 8 & 398.52 & .41 \\
        \makecell[l]{Reconstructed\\preference score}
            & AR interfaces & 5 & 83.91 & .14 \\
        \makecell[l]{Reconstructed\\preference score}
            & Voice commands & 4 & 166.75 & .34 \\
        \makecell[l]{Reconstructed\\preference score}
            & Gestures & 8 & 332.35 & .34 \\
        \bottomrule
    \end{tabular}
\end{table}

For the ranking analyses, participant rankings were converted into Borda-style preference scores. For a family with $k$ alternatives, an item ranked in position $r$ received a score of $k+1-r$, so first place received $k$ points and last place received 1 point.% Participants ranked all alternatives within each family.

\begin{table}[!h]
    \centering
    \small
    \caption{Selected Holm-corrected post-hoc contrasts for the 7-point Likert ratings. Preference uses Borda-style reconstruction from rankings. Effect size is the matched-pairs rank-biserial correlation, oriented as item A minus item B.}
    \label{tab:selected-posthoc-contrasts}
    \setlength{\tabcolsep}{3pt}
    \begin{tabular}{lllrrrrrl}
        \toprule
        Outcome & Family & Contrast & A & B & $\Delta$ & $r_{\mathrm{rb}}$ & $p_{\mathrm{Holm}}$ & Sig. \\
        \midrule
        Rating & \acs{ar} & I3 vs. I2 & 5.73 & 5.70 & 0.03 & 0.03 & $1.000$ & No \\
        Rating & \acs{ar} & I3 vs. I6 & 5.73 & 5.27 & 0.46 & 0.36 & $.036$ & Yes \\
        Rating & \acs{ar} & I2 vs. I6 & 5.70 & 5.27 & 0.43 & 0.32 & $.043$ & Yes \\
        Rating & Voice & C4 vs. C1 & 5.27 & 5.20 & 0.06 & 0.14 & $.401$ & No \\
        Rating & Voice & C1 vs. C2 & 5.20 & 5.00 & 0.20 & 0.51 & $.017$ & Yes \\
        Rating & Voice & C4 vs. C3 & 5.27 & 4.30 & 0.97 & 0.91 & $<.001$ & Yes \\
        Rating & Gesture & G8 vs. G2 & 5.87 & 5.54 & 0.33 & 0.69 & $.004$ & Yes \\
        Rating & Gesture & G2 vs. G6 & 5.54 & 5.40 & 0.15 & 0.15 & $.778$ & No \\
        Rating & Gesture & G8 vs. G1 & 5.87 & 3.62 & 2.25 & 0.98 & $<.001$ & Yes \\
        Preference & \acs{ar} & I3 vs. I2 & 4.28 & 4.28 & 0.00 & 0.02 & $1.000$ & No \\
        Preference & Voice & C1 vs. C4 & 4.32 & 3.07 & 1.25 & 0.62 & $<.001$ & Yes \\
        Preference & Voice & C1 vs. C2 & 4.32 & 3.31 & 1.02 & 0.68 & $<.001$ & Yes \\
        Preference & Gesture & G8 vs. G6 & 7.64 & 6.48 & 1.16 & 0.57 & $<.001$ & Yes \\
        Preference & Gesture & G2 vs. G6 & 6.48 & 5.85 & 0.62 & 0.23 & $.163$ & No \\
        Preference & Gesture & G8 vs. G1 & 7.64 & 2.83 & 4.81 & 0.94 & $<.001$ & Yes \\
        \bottomrule
    \end{tabular}
\end{table}
\FloatBarrier

\subsection{Exploratory Sensitivity Checks}
\label{app:sensitivity-checks}
\autoref{tab:exploratory-sensitivity-summary} summarizes exploratory checks for participant-characteristic effects on modality-level composite scores. %These analyses were not used as primary evidence because the sample was unevenly distributed across age, gender, and questionnaire language. 
Age-group differences survived correction, and the significant post-hoc contrasts are shown in \autoref{tab:age-posthoc}.

\begin{table}[!h]
    \centering
    \footnotesize
    \caption{Exploratory participant-characteristic sensitivity checks.}
    \label{tab:exploratory-sensitivity-summary}
    \setlength{\tabcolsep}{3pt}
    \renewcommand{\arraystretch}{1.15}

    \begin{tabular}{llll}
        \toprule
        Factor & Outcomes tested & Test/correction & Result \\
        \midrule

        \makecell[l]{Gender\\(94 male, 30 female)}
        & \makecell[l]{All\\modalities}
        & \makecell[l]{Mann--Whitney U\\with BH-FDR}
        & \makecell[l]{None significant after\\correction; all $p_{\mathrm{FDR}}\ge .615$} \\

        \makecell[l]{Questionnaire language\\(117 PT, 7 EN)}
        & \makecell[l]{All\\modalities}
        & \makecell[l]{Mann--Whitney U\\with BH-FDR}
        & \makecell[l]{None significant after\\correction; all $p_{\mathrm{FDR}}\ge .491$} \\

        \acs{ati} score
        & \makecell[l]{\acs{ati} item\\ratings}
        & \makecell[l]{Spearman\\correlations with\\BH-FDR}
        & \makecell[l]{No item-level association\\survived correction;\\minimum $p_{\mathrm{FDR}}=.855$} \\

        Age group
        & \makecell[l]{All\\modalities}
        & \makecell[l]{Kruskal--Wallis\\with BH-FDR}
        & \makecell[l]{Significant exploratory\\differences for all\\modalities (see \autoref{tab:age-posthoc})} \\

        \bottomrule
    \end{tabular}
\end{table}

\par\medskip
\noindent\begin{minipage}{\columnwidth}
    \centering
    \small
    \captionof{table}{Holm-significant post-hoc age-group contrasts for participant-level composite scores. Age groups with fewer than five participants were excluded from the Kruskal--Wallis and post-hoc tests.}
    \label{tab:age-posthoc}
    \setlength{\tabcolsep}{3pt}
    \begin{tabular}{llrrrr}
        \toprule
        Outcome & Contrast & $n_A$ & $n_B$ & Median A/B & $p_{\mathrm{Holm}}$ \\
        \midrule
        Mean \acs{ar} clarity & 18--24 $>$ 25--34 & 94 & 12 & 5.33 / 4.50 & $.018$ \\
        Mean \acs{ar} clarity & 44--54 $>$ 25--34 & 10 & 12 & 5.83 / 4.50 & $.018$ \\
        Mean voice applicability & 18--24 $>$ 25--34 & 94 & 12 & 5.60 / 3.90 & $.033$ \\
        Mean gesture applicability & 18--24 $>$ 35--44 & 94 & 6 & 5.22 / 2.22 & $.040$ \\
        \bottomrule
    \end{tabular}
\end{minipage}
\par\medskip

\section{Supplementary Interaction-System Benchmark Details}
\label{app:interaction-benchmarks}
This appendix retains the reproducibility details and complete technical results. The metrics characterize the prototype under controlled conditions.

The latency of the HoloLens' microphone audio stream was tested in a controlled experiment. A synthetic 1\,kHz sine wave lasting 50\,ms was produced from a computer's speaker while simultaneously capturing the microphone's audio. Then, we cross-correlated that pattern in the received microphone data to calculate the microphone's latency. %The microphone's latency is set as the time difference between the end of emitting the sound and detecting it in the incoming audio stream.

\begin{figure}[ht]
  \centering
  \includegraphics[width=0.7\linewidth]{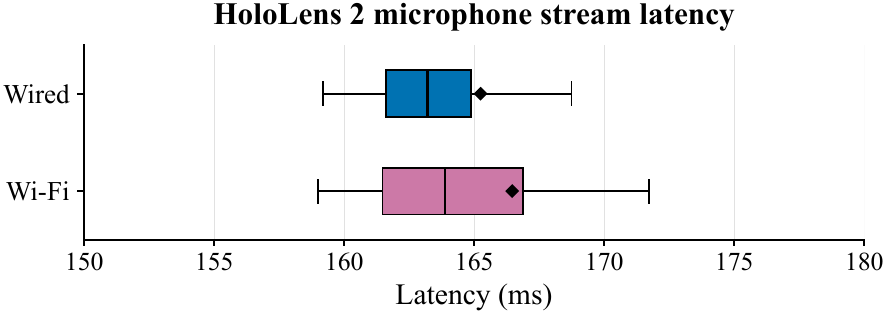}
  \caption{\label{fig:mic}Microsoft HoloLens~2 microphone stream latency over Wi-Fi and wired connections. This control baseline quantifies when speech becomes available to the recognizer.}
  \Description{Microsoft HoloLens~2 microphone stream latency over Wi-Fi and wired connections. This control baseline quantifies when speech becomes available to the recognizer.}
\end{figure}

\autoref{fig:mic} presents the results of the aforementioned test, comparing Wi-Fi to a wired (USB-C) connection. %Box plots show the median and interquartile range; whiskers indicate the 5th-95th percentile range. Values outside this range are not shown for readability.
Wi-Fi conditions provided a clear line-of-sight between the device and router, with more devices on the network under low network load. The results show that both connections achieved similar latencies $M=165.25$\,ms;$SD=11.77$ for wireless and $M=166.46$\,ms;$SD=13.94$ for Wi-Fi. Given the negligible difference, Wi-Fi is used in the remaining evaluation due to its superior mobility and flexibility. %These values serve as a baseline latency for all voice use cases that use the HoloLens' microphone.

To evaluate how computing placement changes the time before a person can transmit intents and receive acknowledgments, we explored two computational scenarios: (1) offloading \ac{asr} processing to a \ac{mec} platform (described in \autoref{subsection:voice-interaction}); and (2) no offloading with processing done on a powerful \ac{vru} compute device. \ac{mec} offloading is recognized as a key enabler for latency-sensitive 5G services within smart cities \cite{taleb2017mec,gholami2022roma}, but it is treated here as one infrastructure-assisted implementation rather than a required architecture.

The \ac{mec} platform hosted a machine with an AMD EPYC 8534P Processor (8 cores) with 16GB of RAM, and an NVIDIA RTX 4000 Ada, operating on Ubuntu 22.04.5 LTS and kernel 6.8.0-52-generic, paired with a Raspberry Pi 5 8GB\footnote{\url{https://www.raspberrypi.com/products/raspberry-pi-5}} as the \ac{vru} compute device. However, when using local processing, tests employed a powerful Jetson Orin Nano Developer Kit\footnote{\url{https://developer.nvidia.com/embedded/jetson-nano}}, featuring a Cortex-A78AE (6 cores) with 8GB of RAM (shared with the GPU), operating on Ubuntu 22.04.5 LTS, kernel 5.15.136-tegra and JetPack 6.0.

Different Whisper \ac{asr} models (Tiny, Base, Small, Medium, and Turbo) were evaluated across these configurations to quantify their impact on system latency. 
%This test involved generating an audio file from the dataset previously fed into Rasa's \ac{nlu}. 
A \ac{tts} tool\footnote{\url{https://ttsmaker.com}} converted all voice commands input into Rasa's \ac{nlu}.% spaced 5 seconds apart, ensuring no overall transcription across commands. Consequently, the command-oriented \ac{nlu} condition should be interpreted as a controlled benchmark of the speech-intent pipeline.

\begin{figure}[ht]
  \centering
  \includegraphics[width=0.7\linewidth]{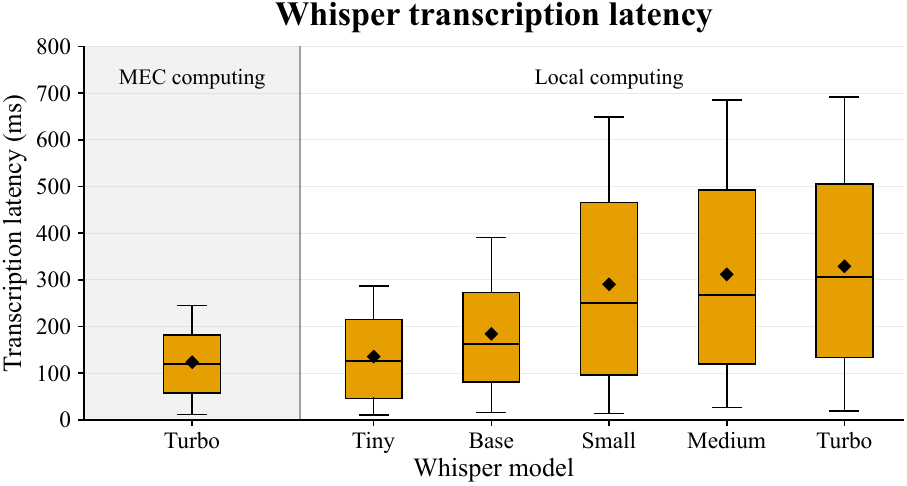}
  \caption{Whisper transcription latency across different models with \acs{mec} offloading and local computing. Lower and less variable values imply more predictable timing for the user to receive recognition feedback.}
  \Description{Whisper transcription latency across different models with \acs{mec} offloading and local computing. Lower and less variable values imply more predictable timing for the user to receive recognition feedback.}
  \label{fig:models-latency}
\end{figure}

\autoref{fig:models-latency} presents the transcription latency of different Whisper \ac{asr} models across local and offload computing to the \ac{mec}.
%The results show that smaller local models reduce transcription latency. Notably, 
The Turbo model on the \ac{vm} configuration achieved a latency ($M=123.74$\,ms;$SD=76.55$). On the Jetson Orin Nano the 
%configurations, we observed incremental increases in latency correlating with the complexity of the model's size: 
Tiny model achieved ($M=135.75$\,ms;$SD=95.10$), Base ($M=184.26$\,ms;$SD=125.21$), Small ($M=290.50$\,ms;$SD=214.91$), Medium ($M=311.69$ \,ms;$SD=216.96$), and Turbo ($M=329.08$\,ms;$SD=220.42$). %Models beyond base impose high mean and standard deviation transcription latency values.
The Turbo model latency was unexpectedly high due to VRAM limitations on the Jetson.% hardware, indicating that hardware constraints significantly impact the model's performance.

\begin{table}[t]
\centering
\small
\setlength{\tabcolsep}{3.5pt}
\begin{tabular}{@{}llccccc@{}}
\hline
\noalign{\vskip 3.5pt}
Hardware &
Model &
\shortstack{Audio\\Condition} &
\acs{wer} (\%) &
\acs{cer} (\%) &
\shortstack{Wake Word\\Detection (\%)} &
\shortstack{Wake Word\\FP (\%)} \\[2pt]
\hline
\acs{vm} & Turbo & \acs{nlu} & \textbf{2.9} & \textbf{1.6} & 96.77 & 0.00\\
Jetson & Tiny & \acs{nlu} & 10.2 & 6.9 & 93.55 & 0.00 \\
Jetson & Base & \acs{nlu} & 7.0 & 5.5 & 93.55 & 0.00 \\
Jetson & Small & \acs{nlu} & 6.6 & 4.5 & 87.10 & 3.74 \\
Jetson & Medium & \acs{nlu} & 3.8 & 3.3 & 96.77 & 0.00 \\
Jetson & Turbo & \acs{nlu} & 4.9 & 3.8 & 90.32 & 0.00 \\
\hline
\acs{vm} & Turbo & \acs{nlu} + Noise & \textbf{3.4} & \textbf{2.0} & 93.55 & 0.00 \\
Jetson & Base & \acs{nlu} + Noise & 8.0 & 6.7 & 87.10 & 0.00\\
\hline
\acs{vm} & Turbo & \acs{fleurs} & \textbf{14.7} & \textbf{10.4} & N/A & N/A \\
Jetson & Base & \acs{fleurs} & 17.3 & 11.9 & N/A & N/A \\
\hline
\end{tabular}

\vspace{.5em}
\footnotesize Bold values indicate the lowest \acs{wer}/\acs{cer} within each audio condition. Wake word metrics are reported only for the \acs{nlu} audio condition.

\caption{Speech recognition \acs{wer}, \acs{cer}, wake word detection, and FP (when applicable), using WhisperLive, organized by hardware, model, and audio condition. Higher error rates indicate a higher likelihood of misclassification of intents.}
\label{tab:metrics_summary}
\end{table}

%The representative \ac{e2e} decomposition \revision{presented in \autoref{fig:e2e-latency} compares the \ac{mec} scenario with the Base local model.} We focus on the Base model instead of the Tiny model, since the lighter model is only recommended for low-powered edge devices due to its low recognition accuracy and higher likelihood of hallucinations. The latency mic stream is the same across configurations, matching the Wi-Fi condition's results. Rasa intent recognition latency is also constant, achieving $M=14.90$\,ms;$SD=3.73$. Although \ac{mec} Turbo had the lowest transcription latency, $M=123.74$\,ms;$SD=76.55$, the offloaded pipeline's communication component dominated the total latency, accounting for 301.39\,ms of the \ac{e2e}'s 440.04\,ms. Local Base avoided this offloading overhead and achieved a lower total latency of 365.63\,ms.

%To evaluate the system's transcription accuracy, 
We computed the \ac{wer} and \ac{cer} on different configurations with three audio conditions: the \ac{nlu} audio file, an \ac{nlu} audio file overlapped with urban background noise~\cite{urbansound}, and the \ac{fleurs} test dataset~\cite{fleurs}. The urban background noise track 
%is a concatenation of clips from the UrbanSound8K test dataset \cite{urbansound} with equal label representation. The 
audio clips were normalized in a round-robin manner for signal-to-noise ratio values of 15, 10, 5, 0, and -5\,dB. Voice activation detection~\cite{SileroVAD} was enabled in Whisper.
%This method ensured varied noise scenarios, essential for the system's transcription accuracy. Regarding noise, it is important to clarify that our evaluation employed the default \ac{vad} \cite{SileroVAD} feature set in WhisperLive. Tests conducted without \ac{vad} yielded significantly worse accuracy, especially in noisy conditions.

%The accuracy evaluation used a separate playback-and-capture setup from the latency measurements. A laptop played the audio files with a Microsoft HoloLens~2, capturing the audio through its microphones in an enclosed room. The computer and the headset were on the same Wi-Fi network, and the laptop processed the audio stream and offloaded the data to the appropriate device, according to the test scenario. 
We followed a similar methodology and the same physical setup as the microphone latency experiment. \autoref{tab:metrics_summary} presents these accuracy measurements. %This setup was used only to compute speech recognition metrics; the \ac{e2e} latency measurements used the Raspberry Pi~5 and 5G connectivity.

The Turbo model in the \ac{vm} configuration achieved the lowest error rates across all scenarios. On the command-oriented \ac{nlu} audio, it obtained \ac{wer} of 2.9\% and \ac{cer} of 1.6\%, while also detecting the wake word ``car'' in 96.77\% of cases with no false positives. %Among local Jetson configurations, larger models generally improved transcription accuracy, but with higher latency cost (see \autoref{fig:models-latency}).

Wake word detection was included because the system should not treat every nearby utterance as a command to the vehicle. In our command grammar, the wake word acted as an activation gate before intent classification. We therefore report both wake word detection, i.e., the percentage of command utterances in which the wake word was correctly detected, and wake word false positives (FP), i.e., the percentage of non-wake or incorrectly segmented utterances in which the system incorrectly detected the wake word. %The \ac{fleurs} dataset was not evaluated for wake word detection because it is a general dataset that does not follow our command grammar.

The Turbo model for the \ac{vm} configuration remained accurate under added urban noise, increasing \ac{wer} to 3.4\% and \ac{cer} to 2.0\%. In contrast, the Base model on the Jetson reached a \ac{wer} of 8.0\% and a \ac{cer} of 6.7\%. The \ac{fleurs} benchmark produced higher error rates for both configurations, which is expected because it contains unconstrained speech rather than the short command grammar. %Nevertheless, the Turbo model on the \ac{vm} configuration outperformed the Base model locally by 2.6\% on \ac{wer} and 1.5\% on \ac{cer}.

These accuracy results come from controlled dataset- and \ac{nlu}-oriented tests rather than from speech collected during situated road-crossing interactions. Nevertheless, a deployed system could improve over time by retraining the \ac{nlu} examples and wake word using consented user-generated data. Such adaptation should increase robustness for recurring users.

\end{document}